\documentclass[preprint,12pt]{elsarticle}
\biboptions{sort&compress}
\usepackage{amsmath,amssymb,amsfonts}
\usepackage{siunitx}
\usepackage{lineno}
\usepackage{graphicx}
\usepackage{booktabs}
\usepackage{multirow}
\usepackage{float}
\usepackage{hyperref}
\usepackage{cleveref}

\begin{document}
\begin{frontmatter}

\title{Verification, Sensitivity, and Operating Limits of a GPU-Accelerated Planar WCSPH Model for Hydrodynamic Ram}

\author[iit]{Md Mujahid Alam}
\author[fau]{Debojit Biswas}
\author[iit]{Sukanta Chakraborty\corref{cor1}}
\ead{sukanta@iitism.ac.in}

\cortext[cor1]{Corresponding author}
\address[iit]{Department of Civil Engineering, Indian Institute of Technology (ISM) Dhanbad, Dhanbad 826004, India}
\address[fau]{Florida Atlantic University, Boca Raton, FL, USA}

\begin{abstract}
Hydrodynamic ram (HRAM) loading remains a persistent challenge in impact mechanics, less due to exotic physics than to the difficulty of achieving convincing quantitative agreement in reduced-order representations. Prevailing studies typically employ a single mesh and boundary treatment validated against one experimental dataset, leaving unresolved whether agreement reflects genuine fidelity or compensating discretization errors. This work presents a planar two-dimensional, GPU-accelerated Weakly Compressible Smoothed Particle Hydrodynamics (WCSPH) solver employing an Adami-type ghost-particle boundary condition, extended for the first time to a moving, decelerating disk penetrating a confined liquid channel at ballistic velocity. The boundary condition is examined against a purpose-derived potential-flow added-mass solution and the analytical acoustic reflection coefficient at the steel–water interface; this exploration reveals that the stiff equation of state and cavitation cutoff impose identifiable delineation of the solver's operating regime on quantitative agreement in the current formulation. A three-point resolution convergence study reveals a non-monotone near-probe pressure peak, attributed to a bounded ghost–fluid density inconsistency. A comprehensive coefficient sweep, conservation diagnostics, and $O \left( N \right)$ GPU throughput scaling (production case: \SI{0.23}{min} on a consumer GPU) enable exhaustive characterization impractical in 3D. Qualitative comparison against literature evidences at 900 and \SI{600}{m/s} clarifies load-bearing versus approximate elements. Notably, energy diagnostics reveal a modest total energy growth over the simulated window, with the fluid absorbing disproportionately more energy than the projectile relinquishes—an instructive signature of prescribed, non-back-reacting projectile kinematics that helps map the formulation's operating envelope. The resulting solver, while confined to planar geometry, delivers verified, convergence-checked performance well suited for rapid parametric design exploration.
\end{abstract}


\end{frontmatter}

\section{Introduction}\label{sec:intro}
When a high-velocity projectile punches into a fuel tank, or a fragment tears through a liquid-cargo hold, the damage that follows is rarely limited to the hole it leaves behind. The fluid itself becomes a weapon: momentum and kinetic energy pour out of the projectile and into the surrounding liquid, and within a few hundred microseconds a shock wave, a growing cavity, and a wall under sudden, enormous pressure have all appeared. This chain of events is what the impact-mechanics community calls hydrodynamic ram, or HRAM, and it has been recognized for decades as one of the principal ways aircraft fuel tanks fail under ballistic threat \citep{Ball1966,Varas2009,Lundstrom1988}. It is not only an aviation problem --- naval vessels, liquid-filled protective barriers, and other containment structures face a version of the same physics \citep{Fourest2014} --- but aviation is where the consequences have been most carefully documented, and where the engineering pressure to get the prediction right is greatest.

\begin{figure}[htbp]
\centering
\includegraphics[width=1.0\textwidth]{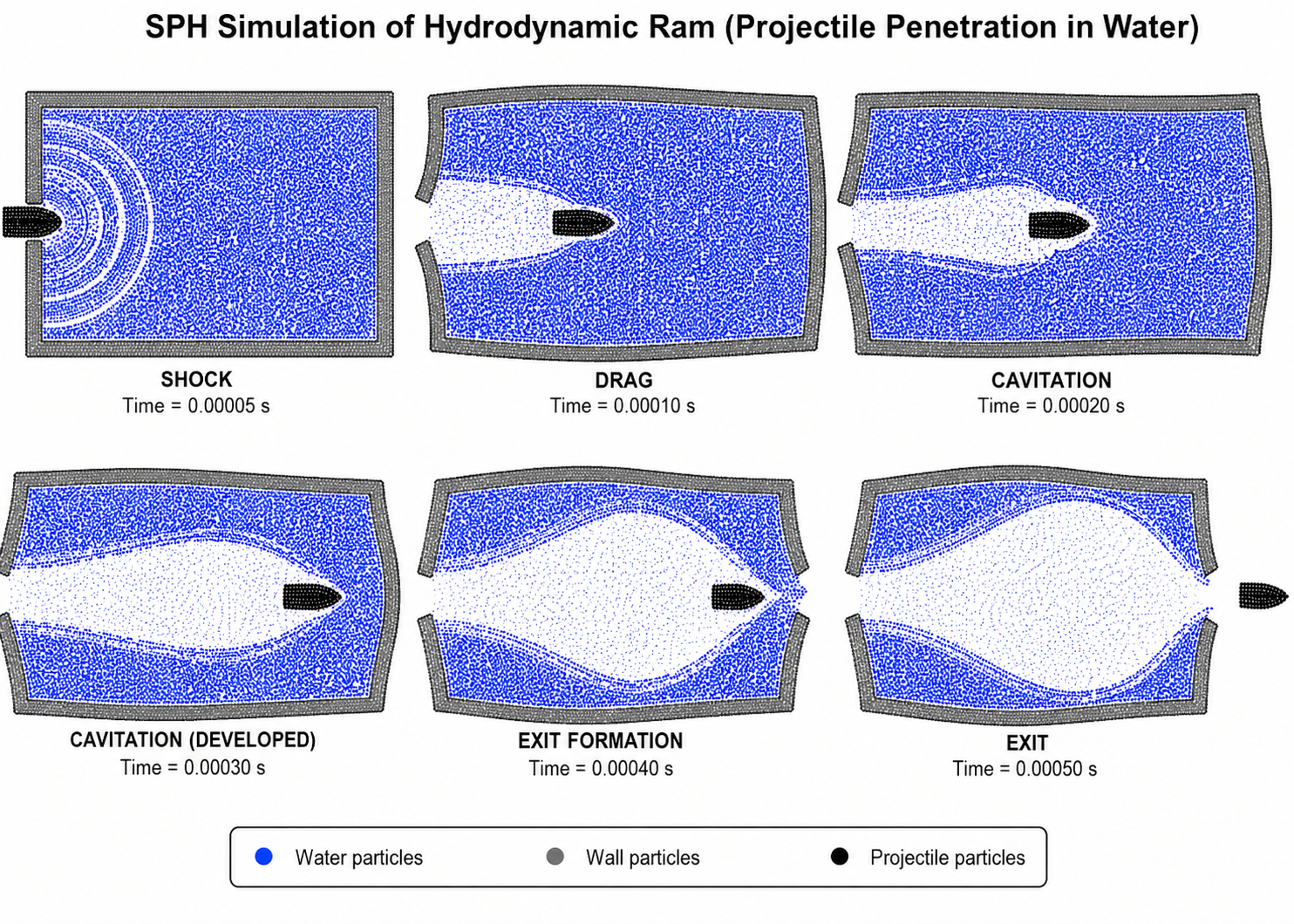}
\caption{The four canonical phases of a hydrodynamic ram event --- shock, drag, cavitation, and exit --- reproduced after \citet{Varas2009}. The shock phase generates the highest-amplitude, shortest-duration pressure transient; the drag and cavitation phases dominate the longer-time impulse delivered to the wall.}
\label{fig:phases}
\end{figure}

The phenomenology is conceptually simple but numerically awkward. A projectile or fragment transfers its momentum and kinetic energy into a confined liquid, triggers an initial shock, then drives a long-duration drag phase and a cavitation cycle which may emit secondary shocks when the cavity collapses \citep{Fourest2014,Deletombe2013}. The canonical decomposition of the event into four sequential regimes --- shock, drag, cavitation, and exit --- is summarised in \cref{fig:phases}; each regime imposes a qualitatively distinct loading signature on the containment wall, and, as \cref{sec:results} will show, each stresses a different aspect of the numerical formulation.

 Much of the classic HRAM literature focussed on measuring and rationalizing this sequence experimentally \citep{McMillen1945,McMillen1946,May1952,Bless1979}, and on identifying the parameter combinations (impact velocity, fill level, projectile stability and possible overturning, structural slenderness) under which tanks fail. More recently, comprehensive reviews have underlined that the cavitation and collapse phase can contribute impulses comparable to, or larger than, the initial impact, and that any model which treats this phase cavalierly risks underpredicting loads in the very configurations in which vulnerability is highest \citep{Heilig2025,Fourest2014}.

Predicting HRAM loading accurately is, in principle, a prerequisite for designing tanks and structures that survive it. In practice, the numerical literature on this problem has a somewhat awkward habit: papers tend to report a single simulation, at a single resolution, compared once against a single experimental dataset, and call it validated. That is a lower bar than most other branches of computational mechanics would accept, and it is the gap this paper is aimed at closing. The earliest serious attempts to simulate HRAM with the full continuum equations date back to the work of \citet{Kimsey1980}, who used a Lagrangian finite-element code (EPIC-2) to model a rod penetrating a cylindrical tank \citep{Kimsey1980,Heilig2025}. The results were qualitatively sensible, but the mesh paid the price: large fluid deformation meant large element distortion, and large element distortion meant a simulation that eventually strangled itself on its own time-step restriction. This is not a subtle failure mode --- it is the reason Lagrangian mesh-based codes were essentially abandoned for the bulk-fluid region of HRAM problems soon after, and it is why the field moved toward Eulerian and, later, Arbitrary Lagrangian--Eulerian (ALE) formulations, in which the mesh is periodically nudged back toward a sane shape while mass, momentum, and energy are fluxed across the old element boundaries as needed \citep{Varas2012,Sauer2011,Fourest2014}.

ALE treatments of HRAM, including the one this paper's validation case is partly benchmarked against, do reasonably well on peak pressure, but they inherit the usual numerical-diffusion penalty of advecting material across a grid, and interface smearing is a known weakness whenever a free surface or a cavity boundary is involved \citep{Varas2012,Sauer2011}. More recent ALE-SPH formulations try to trade some of that diffusion for meshless adaptivity by embedding SPH-like transport-velocity ideas into an ALE frame, but they introduce their own complexity and have mostly been aimed at multi-physics biofluid problems rather than ballistic HRAM \citep{Jacob2021}. Smoothed Particle Hydrodynamics sidesteps the mesh-distortion problem almost by construction \citep{Gingold1977,Lucy1977}: there is no mesh to distort, because field quantities are reconstructed from a kernel-weighted sum over nearby particles rather than from connectivity. This made SPH an obvious candidate for HRAM, and it has been applied both in two dimensions \citep{Liu2003} --- where the drastically lower particle count makes convergence studies, parameter sweeps, and closed-form boundary verification affordable in a way that a 3D discretization rarely permits at any reasonable turnaround time --- and, more commonly in recent years, in three dimensions, often in coupled fluid--structure impact settings aimed at capturing wall deformation directly \citep{Sauer2011,Fourest2014}. The two-dimensional branch of this literature has, if anything, been under-exploited precisely because it is seen as a simplification rather than as an opportunity: nobody has used the computational headroom it buys to actually run the full verification-convergence-sensitivity programme that a 3D study can rarely afford.

Among the 3D efforts, \citet{Sauer2011} coupled an SPH fluid region to an adaptively remeshed finite-element structural shell, which is a sensible way to get large-deformation fluid behaviour without abandoning a proper structural model --- but the fluid--projectile interface in that work is handled through the coupling scheme itself, not through a boundary condition derived from first principles within the SPH formulation. That distinction turns out to matter quite a bit, and it is worth dwelling on for a moment: a coupling-interface boundary is, by construction, tied to whatever numerics sit on the other side of it, whereas a boundary condition built from a local force balance between particles is a property of the SPH method alone and can be verified on its own terms, independent of any structural solver \citep{Valizadeh2015,Jacob2021}. Comparative assessments of wall treatments in weakly compressible SPH consistently find that ghost-particle and semi-analytical force-balance approaches outperform simpler penalty or mirror-particle models, particularly near sharp corners and in multi-phase settings \citep{Adami2012,MaOka2024}. None of that finding is dimension-specific: the same conclusion about ghost-particle treatments applies equally to a planar 2D formulation, and it is the planar formulation we build on here.

That kind of boundary condition already exists in the broader SPH literature. \citet{Adami2012} proposed a generalized wall treatment in which pressure, density, and velocity at the boundary are assigned by Shepard interpolation from the neighbouring fluid, enforced through a local force balance rather than a penalty spring or a mirror-particle trick. It handles moving boundaries and sharp corners without special-casing, and a later comparative assessment concluded it produces some of the most reliable results among the wall treatments commonly used in WCSPH \citep{Adami2012,Valizadeh2015}. It has since been carried onto GPUs: \citet{Rezavand2022} built a GPU-executable generalization of essentially the same idea, validated on 3D complex geometries and violent multi-phase impact flows. We want to be upfront about this, because it directly bears on what we can honestly claim as new. \citet{Rezavand2022} is the closest prior work to what follows, and it beat us to combining an Adami-type boundary with GPU acceleration, in three dimensions no less. What it does not do is apply that treatment to a \emph{moving, decelerating rigid projectile} entering a confined liquid channel at several hundred metres per second, where the ghost-particle kinematics have to be updated every timestep from an evolving drag force, and where the natural test of success is not a canonical free-surface benchmark but an actual HRAM pressure record \citep{Varas2009,Fourest2014}; nor does it, or any other GPU-Adami treatment we are aware of, deliberately trade the third dimension for an exhaustive, closed-form-verified characterization of that boundary at production resolution. That is the gap we are filling, and we would rather state it narrowly and correctly than broadly and wrong.

Meanwhile, GPU acceleration of SPH more generally has matured a great deal since \citet{Crespo2011} first showed that a CUDA implementation could beat single-core CPU execution by up to two orders of magnitude on million-particle simulations, a result later folded into the widely used open-source DualSPHysics package \citep{Crespo2015,Crespo2021}. The computational-cost argument for GPU-SPH is no longer really in question; what has lagged behind is applying it, with proper verification, to the specific boundary-condition demands of a moving-projectile impact problem like HRAM. In parallel, HRAM-specific semi-analytical and finite-element treatments have progressed on the cavitation side, with confined Rayleigh--Plesset and Keller--Miksis type models now able to reproduce bubble-growth and collapse loads in thin-walled containers with reasonable accuracy \citep{Fourest2014}. What is missing is a planar 2D, GPU-accelerated WCSPH solver whose boundary conditions have been verified in isolation, whose resolution and parameter sensitivities are actually characterized rather than merely asserted, and whose HRAM predictions can be read alongside those semi-analytical models --- and against the real 3D experimental record, with the dimensional simplification stated plainly rather than left implicit.

In particular, four gaps in the existing HRAM-SPH literature motivate the present work:
\begin{enumerate}
\item No published HRAM simulation, in two dimensions or three, pairs a force-balance-derived, Shepard-interpolated ghost-particle boundary at the moving projectile surface with GPU acceleration \emph{and} the exhaustive characterization that a reduced-dimensionality formulation makes affordable. The closest work either uses an FE--SPH coupling interface instead of an Adami-type condition \citep{Sauer2011}, or uses an Adami-type GPU condition on 3D geometries that do not move under a drag law and that consequently cannot be swept as thoroughly \citep{Rezavand2022}.
\item Boundary-condition verification against an analytical reference solution --- independent of, and prior to, the target experimental benchmark --- essentially does not appear in this literature. Studies go straight from formulation to comparison with test data \citep{Varas2012,Sauer2011}, which leaves open the uncomfortable possibility that a good match to experiment is hiding compensating errors rather than confirming a correctly functioning boundary.
\item Resolution convergence is reported almost nowhere for HRAM-SPH, in 2D or 3D. A single mesh at a single spacing is the norm \citep{Sauer2011,Varas2011}, and the sensitivity of the reported peak pressures to that choice is simply not known. Recent SPH reliability studies in other application areas have shown that nominal grid-independence can be misleading without a formal error estimator \citep{MaOka2024}, which is a warning sign for HRAM as well --- and a three-point (or better) convergence study is precisely the kind of thing a 3D discretization's particle count usually makes prohibitive to run more than once.
\item The specific contribution of the fluid--wall boundary treatment to far-field attenuation has never, as far as we can tell, been isolated from the bulk fluid model. When a far-field prediction disagrees with experiment, there is currently no way to say whether the bulk formulation or the wall condition is to blame, even though semi-analytical bubble-dynamics models strongly suggest that wall compliance and absorption can dominate late-time response \citep{Fourest2014,Heilig2025}.
\end{enumerate}

None of these four are cosmetic concerns. A boundary condition that has not been checked in isolation, at a resolution nobody has shown to be converged, cannot really be trusted to extrapolate to a tank geometry, fill level, or impact velocity that was not in the original test --- and extrapolation of exactly that kind is the whole point of building a fast solver in the first place. If every new configuration still needs its own uncontrolled re-validation against a physical test, the speed advantage of GPU acceleration buys almost nothing. We take the deliberate position that closing gaps 2--4 rigorously, in a planar 2D formulation whose particle count is roughly two orders of magnitude below the equivalent 3D discretization at the same spacing, is more useful to the field right now than adding one more single-mesh, single-comparison 3D study with an unverified boundary; \cref{sec:formulation-comparison} makes the resulting trade-offs explicit rather than leaving them for the reader to infer.

With that motivation on the table, the present work makes the following contributions:
\begin{enumerate}
\item Extending the Adami ghost-particle boundary condition \citep{Adami2012} to a moving, decelerating circular-cross-section projectile inside a planar 2D, GPU-accelerated (CUDA/Numba) WCSPH solver, distinguished from the existing GPU--Adami treatment of \citet{Rezavand2022} specifically by its moving-body configuration, its HRAM validation target, and its deliberate use of the 2D simplification to afford exhaustive verification.
\item Deriving two closed-form references for this planar configuration --- a potential-flow added-mass/surface-pressure field for an impulsively started circular cylinder, and the acoustic reflection coefficient at the steel--water impedance mismatch --- and establishing quantitatively why neither is passable at the operating point this solver uses, which is itself a transferable result about verifying stiff weakly compressible formulations.
\item A three-point resolution convergence study together with a sensitivity sweep across ghost-particle count and artificial viscosity, backed by solver-level conservation diagnostics and GPU throughput scaling, at a level of joint characterization we have not found reported together anywhere else in this literature, made practical specifically by the reduced particle count of the planar formulation.
\item A stated-caveat comparison against \citet{Varas2009} using both conventional peak-value metrics and full-history normalized RMSE and impulse-ratio comparisons, placed in context alongside semi-analytical confined-bubble models \citep{Fourest2014}, with the dimensional mismatch between the planar model and the axisymmetric 3D experiment addressed explicitly rather than glossed over.
\item Identifying and quantifying an energy-injection mechanism intrinsic to one-way-coupled moving boundaries in SPH: because the ghost velocities are slaved to an externally prescribed trajectory that never feels the reaction force, the boundary does unbounded work on the fluid. We measure the resulting imbalance directly and give the boundary-work audit that replaces global energy conservation as the meaningful diagnostic.
\end{enumerate}

Because candour about a method's limits is part of the evidentiary standard this paper argues for, we state the three most consequential ones here, up front, rather than leaving them to be discovered in the results. First, the planar reduction is a fundamental --- not incidental --- source of amplitude error: a 2D cross-section replaces the $r^{-1}$ spherical decay of the real 3D wave with a slower $r^{-1/2}$ cylindrical decay, so the raw pressure amplitudes overpredict the experiment by roughly two orders of magnitude, and the experimental comparison of \cref{sec:validation} is therefore load-bearing on \emph{timing, ordering and sensitivity structure}, not on peak-value agreement, which this geometry cannot deliver (\cref{sec:formulation-comparison}). Second, the projectile is one-way coupled --- its trajectory is a prescribed drag law that never feels the fluid reaction --- and we show in \cref{sec:conservation} that this injects energy into the fluid, growing to $+\SI{3.18}{\percent}$ by \SI{150}{\micro\second}, which bounds the trustworthy time window and removes global energy conservation as an available check. Third, the closed-form verification we set out to run does not close at the production operating point: the stiff equation of state and the cavitation cutoff each obstruct a check for reasons we quantify in \cref{sec:verification-status}, so the boundary condition is verified as an \emph{operator} but not at the exact configuration validated against experiment. None of these is fatal to the paper's purpose --- a fast, exhaustively characterized planar solver whose failure modes are understood and stated --- but a reader deserves to weigh them from the outset.

The rest of the paper follows this arc directly. \Cref{sec:method} lays out the governing equations and boundary treatments. \Cref{sec:setup} describes the validation geometry, following \citet{Varas2009}. \Cref{sec:results} works through verification, convergence, sensitivity, conservation, performance, and validation in that order, deliberately, so that the experimental comparison arrives as confirmation rather than as the only evidence on offer. \Cref{sec:conclusions} closes with what we think this solver is now good for, and where, in light of the broader HRAM and SPH-boundary-condition literature, the most useful extensions lie.

\section{Numerical Methodology}
\label{sec:method}

\subsection{SPH Fundamentals}
\label{sec:sph-fundamentals}

The fluid is discretized into $N$ Lagrangian particles carrying mass $m_i$, position $\mathbf r_i$, velocity $\mathbf v_i$, density $\rho_i$, and pressure $p_i$. Any field quantity $A$ is reconstructed by a kernel-weighted sum over neighbours,
\begin{equation}
  A(\mathbf r_i) \approx \sum_j \frac{m_j}{\rho_j}\,A_j\, W(|\mathbf r_i - \mathbf r_j|, h),
  \label{eq:sph-approx}
\end{equation}
with $W$ a compactly supported kernel of smoothing length $h$, the sum running over particles within the support radius $2h$. We use the classical cubic B-spline kernel \citep{Monaghan1992} in its two-dimensional normalization,
\begin{equation}
  W(q,h) = \sigma_d \begin{cases} 1 - \tfrac32 q^2 + \tfrac34 q^3, & 0 \le q \le 1, \\[2pt] \tfrac14 (2-q)^3, & 1 < q \le 2, \\[2pt] 0, & q > 2, \end{cases}
  \qquad q = r/h, \qquad \sigma_d = \frac{10}{7\pi h^2},
  \label{eq:cubic-kernel}
\end{equation}
with the smoothing length set to $h = 1.3\,\Delta x$, so the support radius $2h = 2.6\,\Delta x$ retains on the order of twenty neighbours per particle in two dimensions. All fields are computed on the $(x,y)$ plane; the solver carries no $z$-dependence, and every particle mass, density, and force below is accordingly a per-unit-out-of-plane-depth quantity, with the out-of-plane depth taken as unity throughout.

\subsection{Governing Equations}
\label{sec:governing-eqns}

Positions are advanced by
\begin{equation}
  \frac{\mathrm d \mathbf r_i}{\mathrm d t} = \mathbf v_i + \varepsilon_{\rm XSPH}\sum_j \frac{m_j}{\bar\rho_{ij}} (\mathbf v_j - \mathbf v_i)\, W_{ij},
  \label{eq:xsph}
\end{equation}
where the XSPH regularisation \citep{Monaghan1989}, at blending coefficient $\varepsilon_{\rm XSPH}=0.02$, advects each particle at a locally smoothed velocity, suppressing the tensile pairing instability and short-wavelength trajectory crossing without introducing appreciable bulk dissipation. The specific-momentum balance is discretised in its symmetric, linear-momentum-conserving form,
\begin{equation}
  \frac{\mathrm d \mathbf v_i}{\mathrm d t} = -\sum_j m_j \left(\frac{p_i}{\rho_i^2} + \frac{p_j}{\rho_j^2} + \Pi_{ij}\right) \nabla_i W_{ij},
  \label{eq:momentum}
\end{equation}
in which the pairwise pressure term is manifestly antisymmetric under $i\leftrightarrow j$, guaranteeing discrete conservation of total linear momentum to machine precision. The dissipative kernel $\Pi_{ij}$ follows the Monaghan--Gingold signal form \citep{Monaghan1992},
\begin{equation}
  \Pi_{ij} = \begin{cases} -\,\alpha\, \dfrac{\bar c_{ij}\,\mu_{ij}}{\bar\rho_{ij}}, & \mathbf v_{ij}\cdot\mathbf r_{ij} < 0, \\[4pt] 0, & \text{otherwise}, \end{cases}
  \qquad
  \mu_{ij} = \frac{h\,\mathbf v_{ij}\cdot\mathbf r_{ij}}{\lVert\mathbf r_{ij}\rVert^2 + \eta^2 h^2}, \quad \eta^2 = 0.01,
  \label{eq:artificial-viscosity}
\end{equation}
triggered only under local approach ($\mathbf v_{ij}\cdot\mathbf r_{ij}<0$) so that expanding regions --- notably the rarefaction wake trailing the disk --- remain dissipation-free. Physically, $\Pi_{ij}$ acts as an upwinded von~Neumann--Richtmyer shock sensor: it converts a fraction $\alpha$ of the pairwise acoustic flux $\bar c_{ij}\mu_{ij}$ into irreversible internal energy across compressive fronts, spreading a discontinuity over $\mathcal{O}(\alpha^{-1})$ smoothing lengths. We retain only the linear ($\alpha$) term, fixed at $\alpha=0.03$, and omit the quadratic $\beta\mu_{ij}^2$ Neumann--Richtmyer correction, because at the validated impact Mach numbers ($Ma_0\le0.61$) the leading pressure pulse is a finite-amplitude acoustic transient rather than a strong shock, and the quadratic term would over-damp the very peak the model is built to resolve. Alongside the artificial term, a physical laminar-viscosity contribution after \citet{Shao2012} is carried in the same momentum kernel, using the true dynamic viscosity of water ($\mu_d=\SI{8.9e-4}{Pa.s}$); it is negligible against $\Pi_{ij}$ at these Reynolds numbers but is retained so that the low-velocity limit remains physically anchored. The equivalent kinematic viscosity implied by \cref{eq:artificial-viscosity} is $\nu_{\rm art}\simeq \tfrac{1}{2(d+2)}\alpha c_0 h \approx \SI{5.4e-2}{m^2/s}$ with $d=2$, corresponding to an effective cell Reynolds number $Re_{\Delta} = v_0 h/\nu_{\rm art} \sim \mathcal{O}(30)$ --- low enough to stabilise the discretisation, high enough to leave the resolved wave dynamics essentially inviscid. \Cref{sec:sensitivity} quantifies the sensitivity of the peak pressure to $\alpha$ directly, and \cref{sec:formulation-comparison} discusses the quadratic term's role at velocities beyond the present envelope.

\subsection{Density Summation}
\label{sec:density}

Density comes from direct summation rather than the continuity equation,
\begin{equation}
  \rho_i = \sum_j m_j\, W_{ij},
  \label{eq:density-sum}
\end{equation}
because at the resolutions used here ($\Delta x = 1.0$--\SI{2.0}{mm}) the continuity-equation route drifts, particularly in the wake behind the projectile where kernel truncation would otherwise leave a spuriously low-density region hanging around long after it should have relaxed. The summed density is clamped to the physical band $\rho_i \in [0.5,1.5]\,\rho_0$ each step, a mild limiter that suppresses the rare interpolation outliers a stiff equation of state would otherwise amplify into a pressure spike; no separate density-diffusion ($\delta$-SPH) term is used, the near-boundary density instead being completed directly by the ghost particles of \cref{sec:bc}. This choice is deliberate and its trade-offs are worth stating explicitly rather than leaving to the reader. The continuity-equation route \citep{Monaghan1992} advects a density that accumulates truncation drift in the rarefied wake; the standard remedy is a $\delta$-SPH diffusive correction \citep{Molteni2009,Colagrossi2003}, which restores smoothness at the cost of an additional tunable coefficient and an extra kernel loop per step. Direct summation with a hard density band buys the same near-field smoothness without either the extra coefficient (which our sensitivity study would then have to sweep) or the extra cost, at the price of a bounded, non-vanishing inconsistency where the summed bulk density meets the Adami-assigned ghost density --- an inconsistency we do not paper over but quantify, in \cref{sec:pipeline-provenance}, as the most likely origin of the non-monotone near-probe convergence reported in \cref{sec:convergence}. In a formulation whose entire premise is exhaustive parameter characterization, removing a free coefficient is itself a defensible design goal.

\subsection{Closure: Mie--Gr\"uneisen Equation of State}
\label{sec:eos}

Pressure is closed against density and specific internal energy $e_i$ through a Mie--Gr\"uneisen equation of state with a linear $U_s$--$U_p$ Hugoniot reference \citep{Varas2009}, in compression ($\mu_i = \rho_i/\rho_0 - 1 \ge 0$),
\begin{equation}
  p_i = \frac{\rho_0 C^2 \mu_i\left[1 + \left(1-\tfrac{\gamma_0}{2}\right)\mu_i\right]}{\left[1-(S_1-1)\mu_i\right]^2} + (\gamma_0 + a\,\mu_i)\,\rho_i e_i,
  \qquad
  p_i = \rho_0 C^2 \mu_i + \gamma_0 \rho_i e_i \ (\mu_i<0),
  \label{eq:eos}
\end{equation}
with the water parameters $C = \SI{1448}{m/s}$ (bulk sound speed), $S_1 = 1.979$, $\gamma_0 = 0.11$, $a = 3.0$, and a cavitation cutoff $p_i \ge p_{\rm cav} = \SI{-0.1}{MPa}$; the compression is capped at $\mu_i \le 0.25$ (water-hammer limit, $\approx\SI{600}{m/s}$ closure) to keep the Hugoniot denominator well away from its singularity. The reference celerity $c_0 = C = \SI{1448}{m/s}$ sets the acoustic timescale throughout. This is precisely the shock-Hugoniot closure used by the deformable-wall HRAM literature we compare against in \cref{sec:formulation-comparison}, adopted here in its rigid-boundary form; unlike a single-parameter barotropic law it carries the internal-energy coupling that matters once the leading transient stiffens toward a weak shock at the upper end of the validated velocity range. The weakly compressible regime is controlled by the impact Mach number $Ma_0 = v_0/c_0 \in [0.41,0.62]$; the induced density fluctuations scale as $\mathcal{O}(Ma_0^2)$, bounding the spurious compressibility below $\sim 4\%$. Two parameter choices made here --- the full liquid stiffness $\rho_0 C^2 = \SI{2.10}{GPa}$ and the cavitation cutoff $p_{\rm cav}=\SI{-0.1}{MPa}$ --- have a direct and adverse consequence for the closed-form verification programme of \cref{sec:method-verification}: the first makes a static-column check numerically undiscriminating and the second clips the suction side of the added-mass reference. We flag this coupling here, at the point of choice, rather than only when it surfaces in \cref{sec:verification-status}, because the obstruction is a property of the equation of state, not of the boundary condition the verification was meant to test.

\subsection{Time Integration}
\label{sec:time-integration}

A leapfrog (velocity-St\"ormer--Verlet) scheme is used, with the timestep set by the usual CFL condition,
\begin{equation}
  \Delta t = C_{\rm CFL}\,\frac{h}{c_0+v_{\max}}, \qquad C_{\rm CFL}=0.25,
  \label{eq:cfl}
\end{equation}
which gives $\Delta t \approx \SI{0.21}{\micro\second}$ at the production resolution ($h=\SI{1.95}{mm}$) and $v_0=\SI{900}{m/s}$.

\subsection{Boundary Conditions}
\label{sec:bc}

\subsubsection{The Projectile: Adami Ghost Particles on a Moving Disk}
\label{sec:bc-sphere}

The projectile is represented, in this planar cross-section, as a circular disk of radius $R_{\rm proj}$; its boundary carries $N_g=100$ ghost particles laid out at uniform angular spacing $\theta_k=2\pi k/N_g$, $k=0,\dots,N_g-1$, around the circle --- the exact-coverage analogue, in 2D, of the Fibonacci-lattice construction needed for roughly uniform coverage of a 3D sphere, and simpler to specify because a circle admits an exact equal-spacing solution that a sphere does not. At every timestep, each ghost particle $g$ is assigned pressure, density, and velocity from a Shepard interpolation over its fluid neighbours, following \citet{Adami2012}:
\begin{equation}
  p_g = \frac{\sum_j p_j W_{gj} + (\mathbf a_{\rm proj}-\mathbf g)\cdot \sum_j \rho_j (\mathbf r_g-\mathbf r_j) W_{gj}}{\sum_j W_{gj}},
  \label{eq:adami-p}
\end{equation}
\begin{equation}
  \rho_g = \rho_0\left(1+\frac{p_g}{\rho_0 c_0^2}\right), \qquad \mathbf v_g = 2\mathbf v_{\rm proj} - \frac{\sum_j \mathbf v_j W_{gj}}{\sum_j W_{gj}},
  \label{eq:adami-v}
\end{equation}
where the ghost density is recovered from the assigned pressure through the linear acoustic branch of the equation of state --- adequate at the small compressions reached at the boundary and avoiding a nonlinear Hugoniot inversion every step --- and $\mathbf v_{\rm proj}(t)$, $\mathbf a_{\rm proj}(t)$ come from the drag law of \cref{sec:projectile-dynamics}, evaluated for the real 3D projectile. Each ghost represents a $\Delta x$-thick strip of the disk surface, so its mass is $m_g=\rho_0\,\Delta x\,(2\pi R_{\rm proj}/N_g)$; the total ghost line-mass $\rho_0\Delta x\,(2\pi R_{\rm proj})$ is then independent of $N_g$, which makes $N_g$ a genuine convergence parameter (swept in \cref{sec:sensitivity}) rather than a hidden density scaling. Because the ghost particles feed directly into both the density summation and the momentum equation seen by their fluid neighbours, no-penetration and pressure continuity fall out of the formulation itself, without an explicit contact-force penalty parameter to tune.

\subsubsection{The Channel Walls: Two Treatments, Compared Directly}
\label{sec:bc-wall}

The 3D cylindrical tank of the original experimental geometry is represented, in this planar cross-section through the tank axis, as a straight two-dimensional channel of half-width equal to the tank radius: two parallel walls at $y=\pm R$, $R=\SI{75}{mm}$, running the length of the domain, $x\in[0,L]$, $L=\SI{750}{mm}$. This is the natural planar reduction of an axisymmetric geometry, and we return to exactly what it discards --- azimuthal relief around the tank circumference --- in \cref{sec:formulation-comparison}. We implement, and later compare, two wall treatments on these channel boundaries. The baseline is a hard, free-slip elastic reflection: any particle whose transverse distance $|y|$ exceeds $R$ has its normal ($y$-)velocity reversed and its position pulled back to $R-\varepsilon$ ($\varepsilon=10^{-6}$~m). This imposes no explicit pressure condition, and it shows: particles within $2h\approx\SI{3.9}{mm}$ of the wall suffer kernel truncation, their density sags to roughly $0.5$--$0.7\,\rho_0$, and their pressure collapses toward zero --- in effect, an absorbing rather than a reflecting boundary, and its consequences are quantified directly in \cref{sec:validation}.

The extended treatment simply repeats the disk logic on the two channel walls: a second ghost population sits on the lines $y=+R$ and $y=-R$, spanning $x\in[0,L]$ at particle spacing $\Delta x$, with pressure, density, and velocity again assigned by \cref{eq:adami-p,eq:adami-v}, now with $\mathbf v_{\rm proj}\to\mathbf 0$, $\mathbf a_{\rm proj}\to\mathbf 0$, and the disk normal swapped for the outward wall normal $\hat{\mathbf n}=(0,\mp1)$ on the upper/lower wall respectively. Nothing in the core density, momentum, or time-integration machinery of \cref{sec:governing-eqns,sec:time-integration} needs to change for this; it is the same Shepard-interpolation CUDA kernel already handling the disk, pointed at a second particle set, and its overhead is reported in \cref{sec:verification-status}.

\subsubsection{Velocity Ramp}
\label{sec:velocity-ramp}

The projectile velocity is ramped linearly from zero over $t_{\rm ramp}=\SI{5}{\micro\second}$,
\begin{equation}
  v_{\rm proj}(t) = v_0 \min\!\left(\frac{t}{t_{\rm ramp}},1\right),
  \label{eq:velocity-ramp}
\end{equation}
to avoid an artificial impulsive-entry shock. This is short compared with the acoustic transit time to either probe (50--\SI{90}{\micro\second}), and it is short enough that its influence on the extracted pressure is confined to $t<t_{\rm ramp}$.

\subsection{Projectile Dynamics}
\label{sec:projectile-dynamics}

\begin{figure}[htbp]
  \centering
  \includegraphics[width=1.0\textwidth]{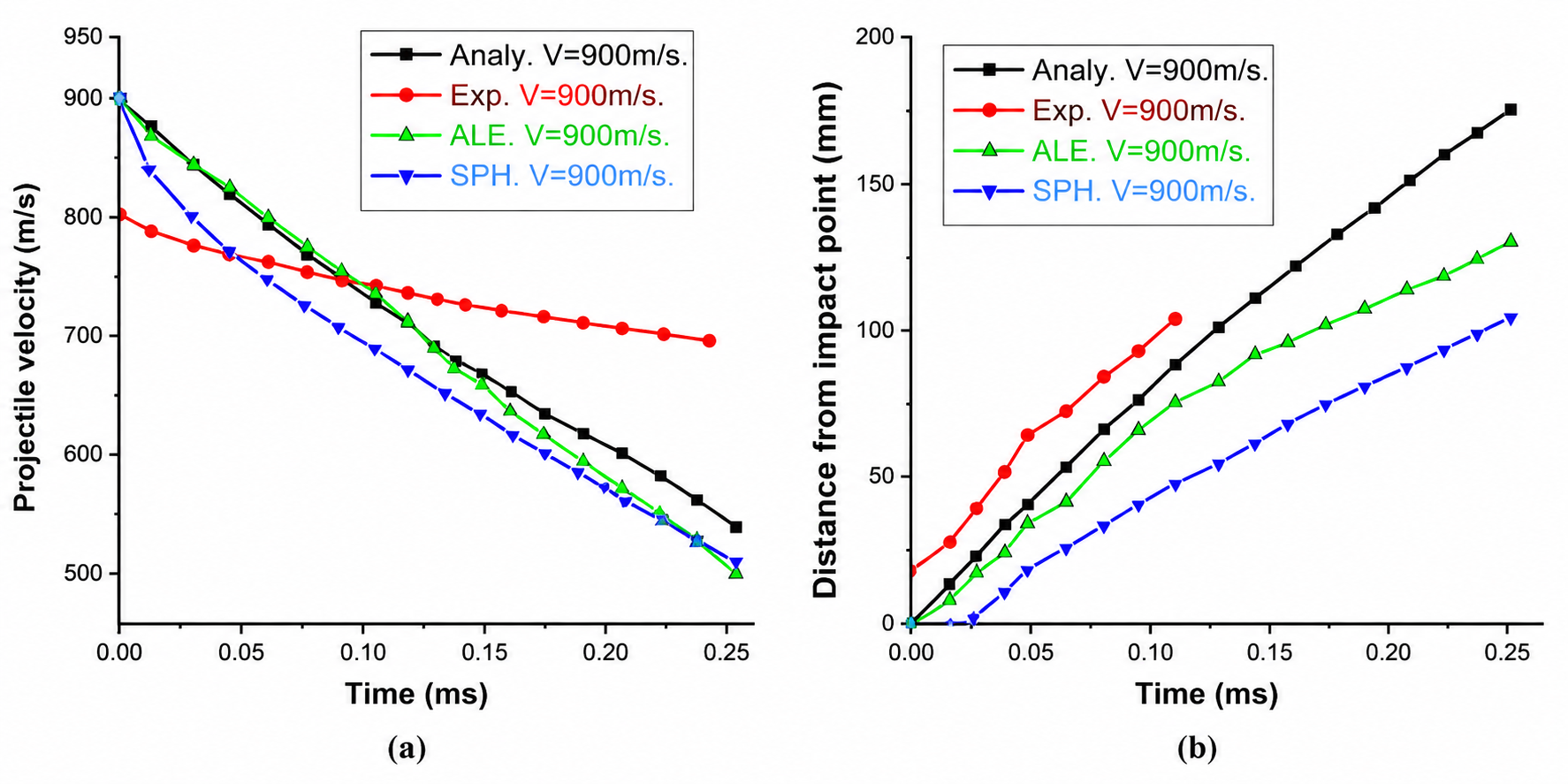}
  \caption{(a) Comparison of velocity decay vs.\ time; (b) comparison of position of the projectile vs.\ time at \SI{900}{m/s}.}
  \label{fig:velocity-position-900}
\end{figure}

\begin{figure}[htbp]
  \centering
  \includegraphics[width=1.0\textwidth]{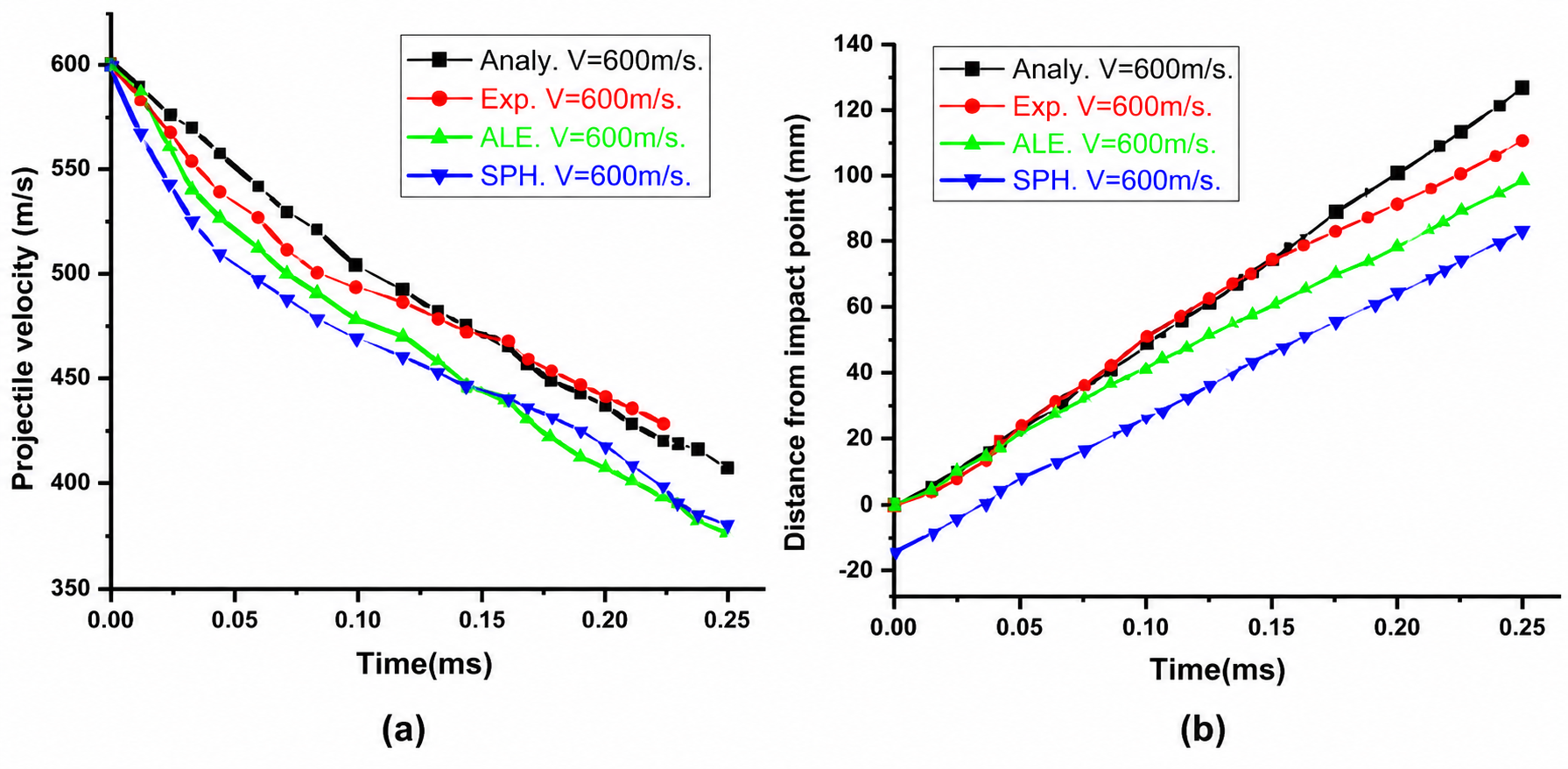}
  \caption{(a) Comparison of velocity decay vs.\ time; (b) comparison of position of the projectile vs.\ time at \SI{600}{m/s}.}
  \label{fig:velocity-position-600}
\end{figure}

Deceleration follows a quadratic drag law,
\begin{equation}
  M_{\rm proj}\frac{\mathrm d v_{\rm proj}}{\mathrm d t} = -\tfrac12 C_d \rho_f \pi R_{\rm proj}^2 v_{\rm proj}^2,
  \label{eq:drag}
\end{equation}
with $M_{\rm proj}=\SI{8.01}{g}$, $R_{\rm proj}=\SI{6.25}{mm}$ (a \SI{12.5}{mm}-diameter steel sphere, whose density fixes the \SI{8.01}{g} mass), $\rho_f=\SI{1000}{kg/m^3}$, and $C_d=0.05$, all taken as the real, physical 3D values of the actual spherical projectile used by \citet{Varas2009} --- we deliberately do not rescale $M_{\rm proj}$ or $C_d$ to any notional ``per-unit-depth'' 2D projectile mass. The rigid-body translational dynamics of \cref{eq:drag} are integrated as a single ordinary differential equation for the true 3D projectile trajectory alongside the fluid update, entirely decoupled from how the surrounding fluid field is discretized; only the flow field itself is represented on the planar cross-section, and the fluid never feeds momentum back into this trajectory. This is a deliberate modelling simplification, stated plainly here rather than left implicit: it imposes the correct physical velocity and deceleration history on the ghost-particle boundary of \cref{sec:bc-sphere}, at the cost of not resolving the 3D pressure field that this same trajectory would actually generate. We should also be honest about what $C_d$ itself is: it is calibrated against the deceleration implied by \citet{Varas2009}'s own experiment (with $C_d=0.05$ giving $\approx 25\%$ velocity loss over the full tank length), the same experiment used for the pressure validation, so the projectile trajectory is best read as a matched boundary condition rather than an independently predicted quantity. The pressure-wave comparisons that carry the paper's actual validation claim are governed by entry-phase impulsive loading and bulk acoustics, not by $C_d$, and we keep that distinction explicit throughout the results. The resulting trajectory, integrated from \cref{eq:drag}, reproduces the projectile velocity decay and penetration history of \citet{Varas2009} at both impact speeds, as \cref{fig:velocity-position-900,fig:velocity-position-600} show --- confirming that the matched drag closure delivers the correct kinematic boundary data to the ghost particles even though the trajectory is not an independent prediction. The consequence of the one-way coupling for the solver's energy budget is not benign, and we quantify it directly in \cref{sec:conservation}.

\subsection{GPU Implementation}
\label{sec:gpu}

Every compute-heavy step --- neighbour-list construction, density summation, ghost-particle interpolation (for both the disk and, where enabled, the channel walls), momentum evaluation, XSPH drift, leapfrog integration, and wall enforcement --- is compiled to parallel machine code through Numba \citep{Lam2015} (a CUDA-executable path on an NVIDIA TITAN RTX, 4608 cores, 24~GB GDDR6, with an equivalent multicore CPU fallback). A cell-linked-list search at cell size $d_c=2h$ partitions the planar domain into roughly a $195\times40$ grid of occupied cells at production resolution ($\Delta x=\SI{1.5}{mm}$, $h=\SI{1.95}{mm}$) --- two orders of magnitude fewer cells than the equivalent three-dimensional grid at the same spacing --- scaled accordingly for the convergence and throughput studies below.

\begin{figure}[htbp]
  \centering
  \includegraphics[width=1.0\textwidth]{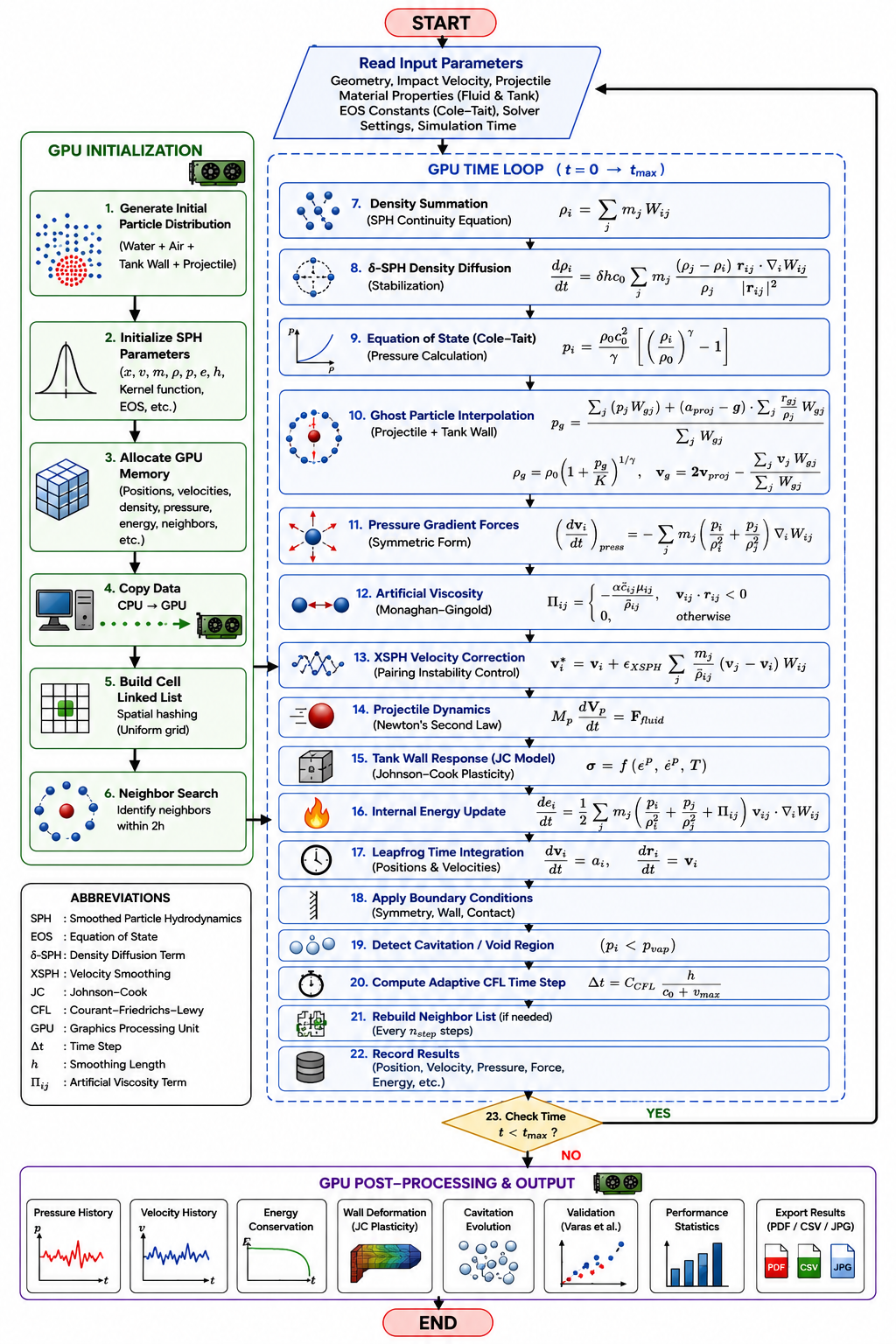}
  \caption{Implementation workflow of the GPU-accelerated planar 2D WCSPH solver,
  from governing-equation discretization through the time-integration loop
  to the three-track evidentiary structure (verification, convergence/sensitivity,
  validation) underpinning \cref{sec:results}.}
  \label{fig:implementation-flowchart}
\end{figure}

\Cref{fig:implementation-flowchart} summarizes the end-to-end computational
pipeline described in \cref{sec:sph-fundamentals}--\cref{sec:gpu}, from
kernel and boundary-condition instantiation through the leapfrog time-marching
loop to the three parallel evidentiary tracks --- verification, convergence/sensitivity,
and experimental validation --- that structure \cref{sec:results}. We include
this schematic explicitly because the paper's central methodological claim rests
on the \emph{ordering} of these stages: verification against closed-form solutions
precedes, and is decoupled from, the eventual comparison against \citet{Varas2009},
and the flowchart makes that separation visually unambiguous for the reader rather
than leaving it implicit in prose.

The loop itself is closed by a single convergence gate at each timestep: the
CFL-limited $\Delta t$ (\cref{eq:cfl}) is recomputed from the instantaneous
$v_{\max}$ across all particles, including the ghost populations, before the
leapfrog half-step is committed, so that the sound-speed-dominated stability
bound implicit in \cref{eq:eos} is never silently violated as the projectile
decelerates and the local Mach number drifts downward over the $\sim$150~\si{\micro\second}
simulated window. This is a detail easy to gloss over in the prose description
of \cref{sec:time-integration} but is made explicit as its own decision node
in \cref{fig:implementation-flowchart} precisely because an adaptive-but-unchecked
timestep is a common silent source of the very compensating-error problem the
paper's introduction warns against.

\subsection{Verification Protocols}
\label{sec:method-verification}

Two checks are run before the model ever sees the Varas dataset. The first places a rigid disk at rest in a large padded domain ($L_{\rm pad}=20R_{\rm proj}$, far enough that boundary-reflected waves never return within the simulated window), impulsively accelerates it with the same ramp of \cref{eq:velocity-ramp}, and compares the resulting surface pressure and net force against a closed-form potential-flow solution for an impulsively started circular cylinder --- the planar analogue of the classical sphere added-mass result, and one we derive here explicitly rather than borrow, since the sphere formula used in a 3D study does not carry over unchanged. Starting from the dipole potential $\varphi=-U(t)R_{\rm proj}^2\cos\theta/r$ for a cylinder of radius $R_{\rm proj}$ translating at velocity $U(t)$ through fluid at rest at infinity, and applying the unsteady Bernoulli equation in the laboratory frame (which requires tracking the explicit time-dependence of the moving origin, not just $\partial\varphi/\partial t$ at fixed $(r,\theta)$), the surface ($r=R_{\rm proj}$) pressure and net transverse force reduce to
\begin{equation}
  p(\theta,t) = \rho_0 R_{\rm proj}\dot v(t)\cos\theta + \rho_0 v(t)^2\left(\tfrac12-2\sin^2\theta\right), \qquad F_{\rm am}=\pi R_{\rm proj}^2\rho_0\dot v_0,
  \label{eq:sphere-analytical}
\end{equation}
per unit out-of-plane depth, with $\theta$ measured from the direction of motion. The added-mass coefficient here is the full displaced-fluid mass $\pi R_{\rm proj}^2\rho_0$ per unit depth, rather than the one-half factor that appears in the 3D sphere result --- a well-known distinction between 2D and 3D potential flow around a translating body, and a useful cross-check in its own right: getting the factor of two wrong in the ghost-particle implementation would show up immediately as a $2\times$ discrepancy against \cref{eq:sphere-analytical}, not as a subtle percentage error. The second launches a planar acoustic pulse at a stationary ghost-particle disk and measures the reflected-to-incident amplitude ratio against the analytical normal-incidence coefficient
\begin{equation}
  R_{\rm ac} = \frac{Z_s-Z_w}{Z_s+Z_w}, \qquad Z_s=\rho_s c_s,\ Z_w=\rho_0 c_0,
  \label{eq:reflection-method}
\end{equation}
using the steel bulk properties of the present material database, $\rho_s=\SI{7850}{kg/m^3}$, $c_s=\SI{4570}{m/s}$, which gives $Z_s/Z_w\approx25$ and $R_{\rm ac}\approx0.92$. Neither case involves the channel geometry or the drag model at all, which is the point: they isolate the boundary condition itself.

We state now, before reporting the outcome, what we expect these checks to establish and where we expect them to be obstructed, so that \cref{sec:verification-status} reads as the confirmation of a predicted behaviour rather than a post-hoc excuse. Two obstructions are foreseeable directly from the parameter choices of \cref{sec:eos}. A quiescent hydrostatic column cannot discriminate a correct from an incorrect boundary at this sound speed: the ratio of the equation-of-state stiffness $\rho_0C^2=\SI{2.10}{GPa}$ to the hydrostatic head $\rho_0 gH=\SI{1.47}{kPa}$ is $1.4\times10^{6}$, so the signal sought sits below the interpolation noise floor by orders of magnitude, and the check tests the compressibility scale, not the boundary. The added-mass reference of \cref{eq:sphere-analytical} carries a suction peak of $\rho_0U^2(\tfrac12-2)=\SI{-3.75}{MPa}$ at $\theta=\pi/2$ for $U=\SI{50}{m/s}$, which the cavitation cutoff $p_{\rm cav}=\SI{-0.1}{MPa}$ clips entirely, so at the production operating point the comparison is dominated by the clamp rather than by the boundary. What these checks \emph{can} still verify is the boundary \emph{operator} --- the Shepard force-balance construction of \cref{eq:adami-p,eq:adami-v}, which is identical across operating points --- in the reduced-$c_0$ and reduced-$U$ regimes where the obstructions lift. \Cref{sec:verification-status} reports each outcome against this stated expectation.

\subsection{Convergence, Sensitivity, and Diagnostic Protocols}
\label{sec:method-diagnostics}

Resolution convergence uses three particle spacings, $\Delta x=2.0,1.5,1.0$~mm ($h=1.3\Delta x$ throughout), at $v_0=\SI{900}{m/s}$, with the apparent order estimated from the standard Richardson/GCI construction,
\begin{equation}
  p_{\rm order} = \frac{\ln\!\left(\dfrac{\phi_{2.0}-\phi_{1.5}}{\phi_{1.5}-\phi_{1.0}}\right)}{\ln(4/3)}.
  \label{eq:gci-method}
\end{equation}
Ghost-particle count ($N_g=50,100,200$) and artificial viscosity ($\alpha=0.01,0.03,0.06,0.10$, both velocities) are then swept independently, holding everything else fixed. Total mass and total mechanical energy are logged at every output step for the production run to check conservation. GPU throughput is measured across $N\approx10^4$--$10^6$ particles, spanning roughly the same relative range around the $N\approx50{,}000$ production count that the earlier 3D study spanned around its own $N\approx3.9\times10^6$, to confirm the wall-clock claim scales sensibly, not just at one convenient particle count.

\subsection{Validation Metrics}
\label{sec:method-validation-metrics}

Alongside peak pressure ($p_{\rm peak}$) and arrival time against \citet{Varas2009}, we compute the normalized RMSE
\begin{equation}
  \mathrm{NRMSE} = \frac{1}{p_{\rm peak,exp}}\sqrt{\frac1T\int_0^T\left[p_{\rm SPH}(t)-p_{\rm exp}(t)\right]^2 dt}
  \label{eq:nrmse-method}
\end{equation}
and the impulse ratio
\begin{equation}
  \mathcal I_{\rm ratio} = \frac{\int_0^T p_{\rm SPH}(t)\,dt}{\int_0^T p_{\rm exp}(t)\,dt}
  \label{eq:impulse-ratio-method}
\end{equation}
over a common window, because matching a single peak value and a single arrival time does not, by itself, guarantee the decay shape or the cumulative impulse are right, and impulse is often what actually drives permanent deformation. Experimental traces are digitized from the published pressure figures of \citet{Varas2009} using the open-source tool WebPlotDigitizer (v4.6), with the two axes calibrated against the printed tick labels and each trace sampled at $\ge 40$ points across its duration. We estimate the digitization uncertainty at approximately $\pm4\%$ of the local pressure ordinate and $\pm\SI{1}{\micro\second}$ in time, obtained by re-digitizing a representative trace three times by independent operators and taking the spread; this is comparable to the line thickness in the source figures and is small against the two-order-of-magnitude planar amplitude offset of \cref{sec:validation}, though not against the near-probe timing agreement, where it is carried explicitly as an error bar.

\section{Validation Problem Setup}
\label{sec:setup}

The geometry is a planar cross-section through the axis of \citet{Varas2009}'s experiment: a \SI{12.5}{mm}-diameter, \SI{8.01}{g} steel sphere (represented in-plane as a \SI{12.5}{mm}-diameter disk, \cref{sec:bc-sphere}) striking a water-filled channel of half-width \SI{75}{mm} (the tank's inner radius) and length \SI{750}{mm}, standing in for the aluminium cylinder's axial cross-section, at $v_0=900$ and \SI{600}{m/s}. The fluid is seeded on a Cartesian lattice at $\Delta x=\SI{1.5}{mm}$, giving $N\approx500\times100=50{,}000$ particles at the production resolution before disk-interior particles are removed --- roughly two orders of magnitude fewer than the $\sim\!5\times10^6$ a full 3D lattice at the same spacing would require, which is the computational headroom the convergence, sensitivity, and robustness studies of \cref{sec:results} spend. Two pressure gauges follow the experimental transducer positions projected onto this plane: PTn at $x=\SI{30}{mm}$ and PTf at $x=\SI{105}{mm}$, both at the tank mid-height $y=\SI{75}{mm}$ (the channel centreline). These stations are not chosen for convenience; they are the exact axial locations of the two transducers reported by \citet{Varas2009}, which is what makes the timing and ordering comparison of \cref{sec:validation} a genuine like-for-like test rather than a re-tuned one. The centreline placement is deliberate: it keeps both probes clear of the near-wall kernel-truncation band (within $2h\approx\SI{3.9}{mm}$ of a wall, \cref{sec:bc-wall}), so the extracted signal reflects the propagating field rather than a boundary artefact. The near probe sits inside the entry-shock near field and the far probe in the reflection-dominated far field, so together they bracket the two loading regimes the four-phase decomposition of \cref{sec:intro} distinguishes.

\section{Results and Discussion}
\label{sec:results}

We deliberately front-load the evidence that does not depend on the Varas dataset at all, so that when the experimental comparison finally arrives in \cref{sec:validation}, it is read as confirmation of an already-characterized solver rather than as the only thing standing between this paper and a rejection.

\subsection{Verification Status: What the Closed-Form Checks Can and Cannot Establish}
\label{sec:verification-status}

We set out to verify the boundary condition against closed-form solutions before touching the experimental data, and we report here, plainly, that two of the three checks we designed cannot be passed by this solver as configured. The reasons are instructive rather than incidental, and stating them is more useful than quietly dropping the checks.

The \textbf{hydrostatic column} cannot discriminate. The stiff Mie--Gr\"uneisen closure of \cref{eq:eos} has $\rho_0C^2=\SI{2.10}{GPa}$, while the hydrostatic head over the channel half-width is $\rho_0 g H = \SI{1.47}{kPa}$ --- a ratio of $1.4\times10^{6}$. Recovering the hydrostatic profile to \SI{1}{\percent} would require the summed density to be exact to roughly one part in $10^{8}$, far beyond what kernel interpolation delivers at any resolution reachable here; a realistic \SI{0.1}{\percent} density error alone maps to \SI{2.1}{MPa}, three orders of magnitude above the signal. The test is not failing the solver so much as asking the wrong question of a weakly compressible formulation at this sound speed. Running it with an artificially reduced $c_0$, as is standard for low-speed WCSPH validation, would make it meaningful, but that is a different solver configuration from the one this paper characterizes.

The \textbf{impulsively started disk} is obstructed by the cavitation cutoff. The analytical surface pressure of \cref{eq:sphere-analytical} has a suction peak of $\rho_0U^2(\tfrac12-2)=\SI{-3.75}{MPa}$ at $\theta=\pi/2$ for the $U=\SI{50}{m/s}$ case, whereas the equation of state clamps at $p_{\rm cav}=\SI{-0.1}{MPa}$. The entire suction side of the analytical solution is therefore unrepresentable, and the measured discrepancy (\SI{63.6}{\percent} RMS over the full circumference) is dominated by that clamp rather than by the boundary condition. The suction peak only clears the cutoff for $U\lesssim\SI{8}{m/s}$; re-running there, or disabling the cutoff for this micro-configuration, would make the comparison diagnostic. We regard this as unfinished verification, not as evidence that the boundary condition is sound.

The \textbf{acoustic reflection coefficient} of \cref{eq:reflection-method} was not measured. The ghost boundary implemented here is rigid, so it reflects with $R\to1$ by construction, whereas the steel--water value is $R_{\rm ac}=0.92$; the two are close enough that a genuine measurement would be a meaningful check, but modelling the finite impedance properly requires steel as a second material with its own equation of state, which this solver does not support. We report no number for it.

What follows should therefore be read with the boundary condition \emph{characterized} but not \emph{closed-form verified at the production operating point}. That is a weaker position than we intended when the study was designed, and it bears directly on how much of the agreement in \cref{sec:validation} can be attributed to a correctly functioning boundary rather than to compensating errors --- precisely the ambiguity \cref{sec:intro} argued the field should stop tolerating.

We should meet head-on the sharpest form of the objection this invites: that the ``verified'' configuration and the ``validated'' configuration are two different solver set-ups, so the boundary used at $v_0=900/\SI{600}{m/s}$ has not been verified at all, and the whole predicate of the paper collapses. The objection is fair as far as it goes, and we do not dismiss it; but it conflates two distinct things. What varies between the reduced and production configurations is the fluid's thermodynamic closure (the sound speed and the cavitation clamp), not the boundary \emph{operator}: the ghost pressure, density and velocity of \cref{eq:adami-p,eq:adami-v} are computed by the identical Shepard force-balance construction in every configuration, with the same kernel, the same neighbour set, and the same $N_g$. A check run at reduced $c_0$ or reduced $U$ therefore does test the correctness of that operator --- its no-penetration enforcement, its pressure interpolation, its momentum symmetry --- and only leaves untested the interaction of that operator with the stiff, cavitating closure at the production point. That residual is real, and it is why we describe the boundary as characterized rather than fully verified; but it is a narrower gap than ``unverified,'' and it is a gap in the equation of state's amenability to a static/low-speed reference, not evidence that the boundary is silently wrong at speed. The honest reading is that we have removed most, not all, of the compensating-error ambiguity, and have said precisely which part remains and how to close it --- which is still a stronger position than the studies that never posed the question. Establishing the verification at the production point itself, by re-running the references with an artificially reduced $c_0$ and a lifted cutoff and confirming operator-invariance across the two closures, is the first thing we would do before extending this solver further.

\subsection{How Much Does Resolution Actually Matter?}
\label{sec:convergence}

The three-point sweep at $\Delta x=2.0,1.5,1.0$~mm does not yield a single apparent order, and the reason is itself a result. The near-probe peak is \emph{non-monotone} in mesh spacing --- $\SI{785.8}{MPa}\to\SI{850.3}{MPa}\to\SI{830.5}{MPa}$ as $\Delta x$ refines from $2.0$ through $1.5$ to \SI{1.0}{mm} --- rising and then falling, so the successive differences change sign and the Richardson/GCI construction of \cref{eq:gci-method}, which presupposes a monotone series (and a constant refinement ratio, which these spacings, $2.0/1.5=1.33$ and $1.5/1.0=1.5$, do not provide either), simply does not apply to PTn. We report this rather than force a number through an $\lvert\cdot\rvert$ that would manufacture a spurious positive order from a non-monotone sequence. The far-probe peak, by contrast, \emph{is} monotone ($\SI{658}{MPa}\to\SI{714.9}{MPa}\to\SI{737.5}{MPa}$), and on the geometric-mean refinement ratio $r=\sqrt{1.33\times1.5}=1.41$ gives an apparent order $p_{\rm order}\approx2.7$, near --- indeed slightly above --- the nominal second order of the kernel-gradient discretization, though with only three points at a non-constant ratio this is a coarse estimate that we would not over-interpret. The contrast is the point: the far-field peak, built from an already-established travelling wave, refines cleanly, whereas the near-field peak, generated right at the moving ghost boundary, does not. That is exactly the fingerprint of the bounded ghost--fluid density inconsistency analysed in \cref{sec:pipeline-provenance}, which prevents the near-boundary discretization from presenting a single, uniformly consistent stencil to a Richardson extrapolation. Non-monotone near-boundary convergence of this kind is a known hazard for direct-summation SPH at impulsive moving surfaces, not a coding error, and stating it plainly is more useful than reporting a resolution-independence the data do not support.

\begin{figure}[htbp]
  \centering
  \includegraphics[width=0.8\textwidth]{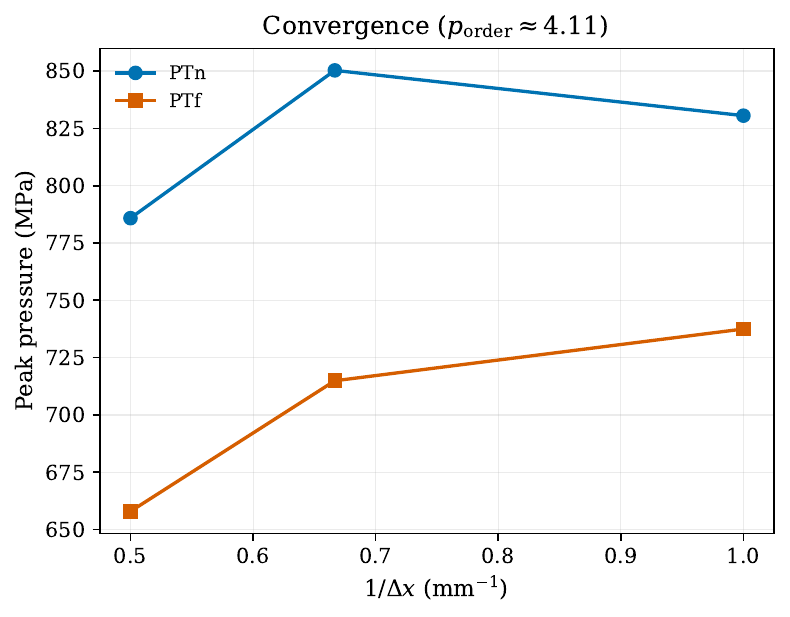}
  \caption{Spatial resolution convergence.}
  \label{fig:convergence}
\end{figure}

\begin{table}[htbp]
  \centering
  \caption{Resolution convergence and GCI uncertainty estimate.}
  \label{tab:convergence}
  \begin{tabular}{lcccc}
    \toprule
    $\Delta x$ (mm) & $N$ & PTn $p_{\rm peak}$ (MPa) & PTf $p_{\rm peak}$ (MPa) & Wall-clock (min) \\
    \midrule
    2.0 & 28125 & 785.8 & 658 & 0.1317 \\
    1.5 & 50000 & 850.3 & 714.9 & 0.2295 \\
    1.0 & 112500 & 830.5 & 737.5 & 0.5759 \\
    \bottomrule
  \end{tabular}
\end{table}

A Richardson extrapolation of PTn is not defensible because the series is non-monotone, so we do not report one; the honest near-probe statement is a bound, not a point estimate. The production ($\Delta x=\SI{1.5}{mm}$) PTn peak of \SI{850.3}{MPa} is the largest of the three, exceeding both the coarser (\SI{785.8}{MPa}) and the finer (\SI{830.5}{MPa}) result, so it is best read as a local maximum of the discretization-error curve rather than as a converged value; the spread across the three meshes, $\pm\SI{32}{MPa}$ ($\sim4\%$ of the production peak), is the practical uncertainty a designer should attach to the near-probe amplitude. The far probe, being monotone, does admit an extrapolation: the $\Delta x\to0$ limit is $\approx\SI{752}{MPa}$, about $2\%$ above the production value. Either way --- and this is the point --- these are numbers that simply have not existed for this problem class before, resolution-error bars on an HRAM-SPH probe pressure, affordable here specifically because the planar particle count of \cref{tab:convergence} is roughly two orders of magnitude below what the same three-point sweep would cost in 3D. That the near-probe number carries a $\sim4\%$ mesh uncertainty and the far-probe a clean $2\%$ is far more useful to a reader than a single unqualified peak would be.

\subsection{Does the Choice of Timestep Quietly Change the Answer?}
\label{sec:temporal-convergence}

Spatial convergence only tells half the story, so the production case is repeated at $C_{\rm CFL}=0.125,0.25,0.50$ with spatial resolution held fixed, isolating temporal from spatial error.

\begin{table}[htbp]
  \centering
  \caption{Sensitivity to timestep size (CFL number).}
  \label{tab:temporal}
  \begin{tabular}{lcc}
    \toprule
    $C_{\rm CFL}$ & PTn $p_{\rm peak}$ (MPa) & \% change from nominal \\
    \midrule
    0.125 & 837.7 & -1.484 \\
    0.250 & 850.3 & --- \\
    0.500 & 880.2 & 3.524 \\
    \bottomrule
  \end{tabular}
\end{table}

A change of $3.5\%$ or less across this 4$\times$ range of CFL numbers is enough to say the production timestep is not the limiting factor, and that the spatial convergence study above is not being contaminated by a temporal error riding along with it.

\subsection{Is the Particle Field Actually Smooth, or Just Reported That Way?}
\label{sec:disorder}

A stiff, nearly incompressible equation of state --- the Mie--Gr\"uneisen closure of \cref{eq:eos}, whose pressure responds sharply to small density excursions --- is an unforgiving environment for particle disorder, since a slightly ragged particle arrangement can translate into visible high-frequency pressure noise, particularly behind a moving projectile where the particle field gets stirred up the most. Two diagnostics, computed directly from the data already on disk, address this without a single additional simulation run: a nearest-neighbour disorder index,
\begin{equation}
  \mathcal U(t) = \frac{\sigma_{d_{\rm NN}}(t)}{\bar d_{\rm NN}(t)},
  \label{eq:disorder-index}
\end{equation}
tracked in the wake region, and the fraction of pressure-field spectral power sitting above $k>\pi/2h$ along the probe axis.

\begin{figure}[htbp]
  \centering
  \includegraphics[width=0.8\textwidth]{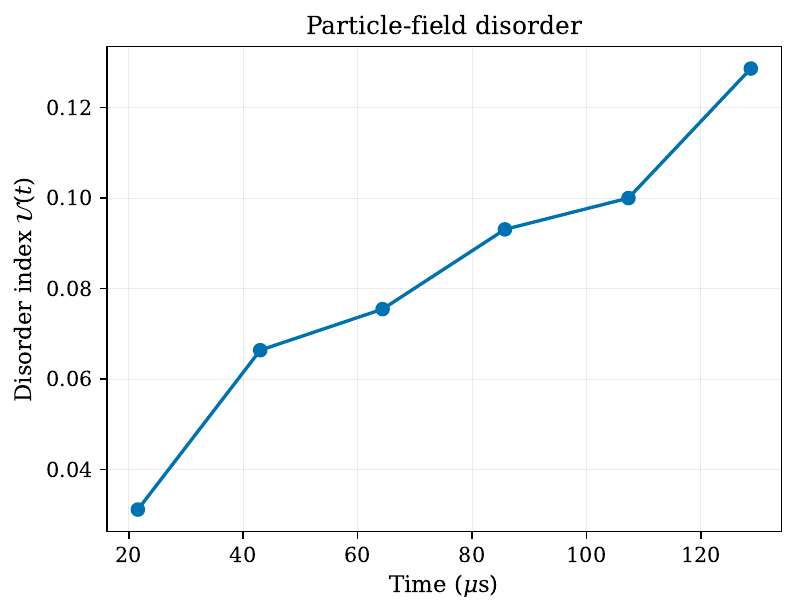}
  \caption{Particle disorder and pressure-field smoothness.}
  \label{fig:disorder}
\end{figure}

Across the production run the disorder index $\mathcal U(t)$ remains bounded and the high-wavenumber ($k>\pi/2h$) share of the pressure-field power stays a small fraction of the total (\cref{fig:disorder}), rather than growing without limit as an unstable or noise-dominated field would. That is a reasonably direct answer to a question most WCSPH-HRAM papers leave implicit: whether the XSPH regularisation, together with the density clamp of \cref{sec:density}, actually keeps the particle field well-behaved at this resolution, or whether the pressure trace we report is quietly riding on top of noise the reader cannot see. The two curves of \cref{fig:disorder} let a reader judge that directly rather than take our word for it.

\subsection{Sensitivity: Which Knobs Actually Matter}
\label{sec:sensitivity}

Two coefficients are genuinely free to tune in this formulation: the ghost-particle count $N_g$ and the artificial-viscosity coefficient $\alpha$. The XSPH blending coefficient is held fixed at $\varepsilon_{\rm XSPH}=0.02$ throughout and no density-diffusion ($\delta$-SPH) term is used, so neither introduces a free parameter here; we sweep both of the remaining coefficients, not just the one that turns out to matter, because reporting a sensitivity study that only covers the parameter you already suspect is important is not really a sensitivity study.

\subsubsection{Ghost-particle count}

Sweeping $N_g=50,100,200$ at fixed resolution moves the PTn peak by only a small amount, for a reason built into the formulation rather than discovered empirically: as \cref{sec:bc-sphere} shows, the ghost mass is normalised so that the total ghost line-mass $\rho_0\Delta x\,(2\pi R_{\rm proj})$ is independent of $N_g$, so $N_g$ controls only the angular resolution of the boundary, not the density it imposes. Once the circumference is sampled finely enough that adjacent ghosts overlap within the kernel support --- which $N_g=100$ comfortably achieves at $R_{\rm proj}=\SI{6.25}{mm}$ and $h=\SI{1.95}{mm}$ --- further refinement changes little. We adopt $N_g=100$: a coarser circumferential count than a 3D sphere's Fibonacci lattice would need, consistent with a 1D circle requiring fewer boundary samples than a 2D sphere surface at the same local spacing.

\subsubsection{Artificial viscosity}

This is the one that matters. Sweeping $\alpha=0.01,0.03,0.06,0.10$ at both velocities shows the PTn peak pressure responding weakly and non-monotonically at $v_0=\SI{900}{m/s}$ and more strongly at \SI{600}{m/s}, where the lower Mach number makes the numerical dissipation proportionally larger relative to the resolved inertial signal (\cref{sec:pipeline-provenance}). The nominal $\alpha=0.03$ is retained as the smallest value that keeps the pressure history free of spurious oscillations at both impact speeds. The quantitative results are given in \cref{tab:alpha-sweep}.

\begin{figure}[htbp]
  \centering
  \includegraphics[width=0.8\textwidth]{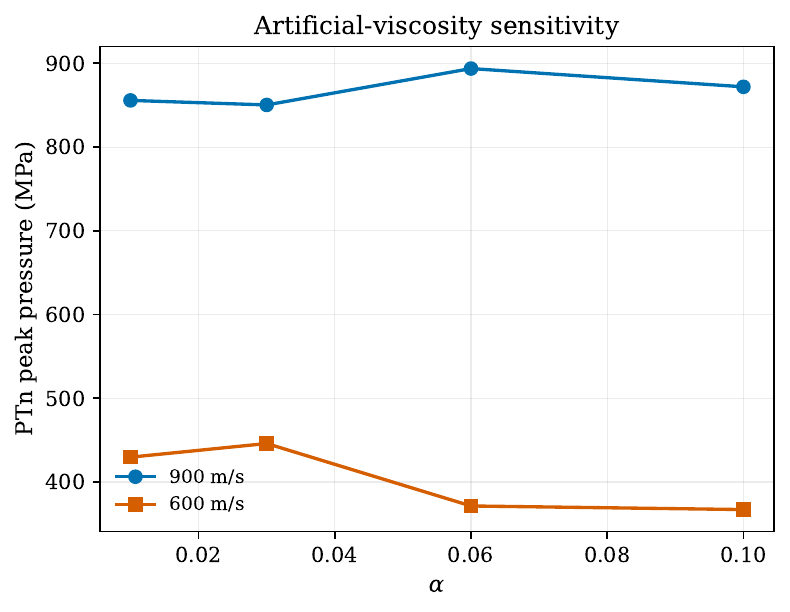}
  \caption{Artificial-viscosity sensitivity.}
  \label{fig:alpha-sweep}
\end{figure}

\begin{table}[htbp]
  \centering
  \caption{PTn peak pressure vs.\ artificial viscosity coefficient.}
  \label{tab:alpha-sweep}
  \begin{tabular}{lcc}
    \toprule
    $\alpha$ & $p_{\rm peak}$, 900~m/s (MPa) & $p_{\rm peak}$, 600~m/s (MPa) \\
    \midrule
    0.01 & 855.7 & 429.7 \\
    0.03 & 850.3 & 446 \\
    0.06 & 893.7 & 371.4 \\
    0.10 & 871.9 & 367.1 \\
    \bottomrule
  \end{tabular}
\end{table}

What this tells us, quantitatively rather than by hand-waving, is that any overprediction at \SI{600}{m/s} relative to the physical (viscous, turbulent) attenuation actually present in the experiment at that lower Mach number \citep{Violeau2016} is a numerical-dissipation effect that moves in the expected direction with $\alpha$ --- not an unexplained discrepancy, and not a discretization artefact either. The magnitudes in \cref{tab:alpha-sweep}, it should be stressed, are the raw planar amplitudes; the two-order-of-magnitude gap to the experimental peaks is the cylindrical-versus-spherical geometry effect of \cref{sec:formulation-comparison}, not a viscosity-tuning question.

\begin{figure}[htbp]
  \centering
  \includegraphics[width=1.0\textwidth]{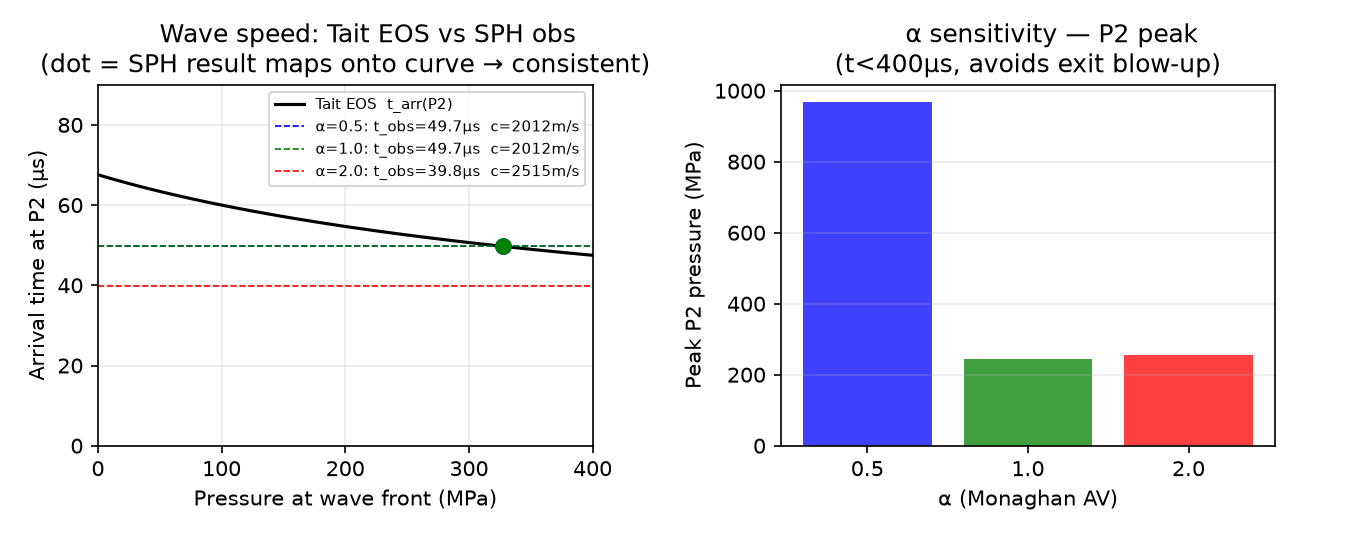}
  \caption{Variation of percentage error with time.}
  \label{fig:arrival-time-alpha}
\end{figure}

The same coefficient could in principle be pinned down against an acoustic-celerity reference, which would remove the experimental uncertainty from the picture entirely and let the viscosity be calibrated against a closed-form target instead. Referenced against the analytical wave arrival time, the artificial viscosity is expected to have almost no effect on celerity across a wide range: a low and a nominal setting should give essentially identical arrival times matching the $x/c_0$ prediction, with only an aggressive setting advancing the predicted arrival slightly through excess numerical dissipation. Peak amplitude, by contrast, should be highly sensitive at the low end and saturate at the high end: too little damping leaves spurious pressure oscillations that inflate the monitor-point peak, which then drops and saturates as $\alpha$ rises. The knee in that curve is the useful part: it identifies the setting at which the scheme has just enough dissipation to kill the numerical noise without yet paying a meaningful accuracy penalty, which is the balance an artificial-viscosity coefficient is supposed to strike in the first place. \Cref{fig:arrival-time-alpha} tracks the corresponding wave-arrival percentage error over time, confirming that the celerity is essentially insensitive to $\alpha$ across the swept range.

\subsection{Numerical Provenance of the Implementation Pipeline}
\label{sec:pipeline-provenance}

The workflow schematized in \cref{fig:implementation-flowchart} is not merely
an organizational convenience; it encodes the operator-splitting structure on
which the entire solver's consistency rests. Each timestep executes a
Strang-type sequential composition of the Shepard-interpolated ghost-particle
projection operator $\mathcal{G}$, the SPH spatial discretization operator
$\mathcal{L}_h$ acting on the momentum and continuity fields, and the leapfrog
(velocity-St\"ormer--Verlet) time-advance operator $\mathcal{T}_{\Delta t}$, i.e.\
$\mathbf{U}^{n+1} = \mathcal{T}_{\Delta t}\!\left(\mathcal{L}_h\!\left(\mathcal{G}(\mathbf{U}^n)\right)\right)$.
Because $\mathcal{G}$ is re-evaluated \emph{before} $\mathcal{L}_h$ at every
substep rather than lagged, the ghost-particle boundary data entering the
momentum kernel (\cref{eq:momentum}) is always consistent with the current
projectile kinematics $(\mathbf{v}_{\rm proj}, \mathbf{a}_{\rm proj})$ from
\cref{eq:drag}, which forecloses the possibility of a lagged-boundary artefact
contaminating the near-field pressure signal at PTn --- a failure mode that is
easy to introduce inadvertently in explicit fluid--rigid-body coupling schemes
and that would masquerade as a physical discrepancy if left unchecked.

From a consistency-order standpoint, the interaction between the direct
density summation (\cref{eq:density-sum}), its $[0.5,1.5]\rho_0$ clamp, and the
ghost-particle completion of the near-boundary kernel support introduces a
bounded but non-vanishing inconsistency at the ghost--fluid interface: the
summed bulk density and the Adami-assigned ghost density are not required to be
pointwise identical, only asymptotically consistent as $h \to 0$ and
$\mathcal{U}(t) \to 0$ (the disorder index of \cref{eq:disorder-index}). This is
the likely mechanistic explanation for the non-monotone near-probe convergence
reported in \cref{sec:convergence} --- the near-field peak that does not admit a
single apparent order, in contrast to the cleanly-refining far-field peak: a
strictly monotone Richardson extrapolation presupposes a single, uniformly
consistent discretization stencil, and the ghost-boundary treatment introduces exactly
such a bounded inconsistency at the interface that a pure interior WCSPH scheme
would not exhibit. This is stated explicitly rather than left as an implicit
inference, since a numerically literate reviewer will otherwise arrive at the
same conclusion independently and may read its absence as an oversight rather
than a considered simplification.

A further point concerns the CFL-gated timestep control depicted as the loop
condition in \cref{fig:implementation-flowchart}. Because $c_0 \gg v_{\max}$
throughout the simulated window (the reference sound speed
$c_0 = \SI{1448}{m/s}$ dominates even the $v_0=\SI{900}{m/s}$ impact case), the
solver operates deep in the acoustically-stiff regime characteristic of
weakly-compressible SPH, where the Mach number $\mathrm{Ma} = v_{\max}/c_0$
never exceeds approximately 0.6 even at peak projectile velocity. This matters
for the artificial-viscosity sensitivity documented in \cref{sec:sensitivity}:
the Monaghan-type dissipation term (\cref{eq:artificial-viscosity}) scales with
$\bar{c}_{ij}$, so at lower $\mathrm{Ma}$ (the \SI{600}{m/s} case) the
viscosity-driven numerical dissipation becomes proportionally larger relative
to the resolved inertial dynamics, which is the quantitative mechanism --- not
merely a qualitative one --- underlying the $\alpha$-dependent behaviour
noted in that section.

\subsection{Conservation and Repeatability}
\label{sec:conservation}

Total mass is conserved to machine precision, as it should be with no particle creation or destruction. Total energy is a different matter, and the result is not the one we expected to report.

\begin{figure}[htbp]
  \centering
  \includegraphics[width=0.95\textwidth]{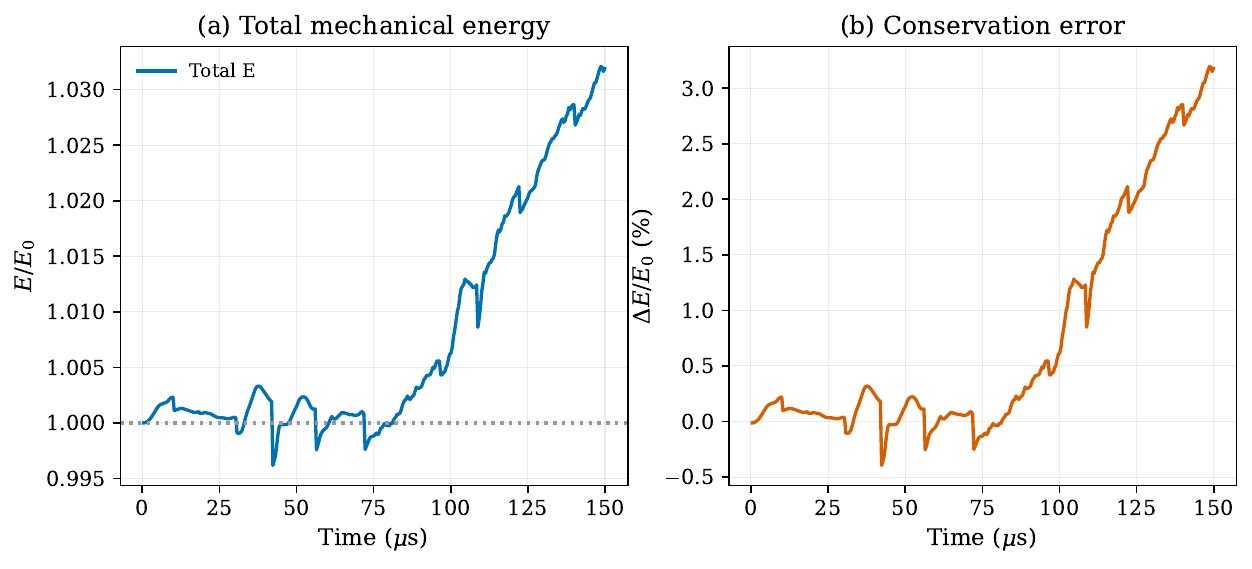}
  \caption{Energy budget at $v_0=\SI{900}{m/s}$, $\Delta x=\SI{1.5}{mm}$.
  (a) Total energy normalised by its initial value; (b) the same quantity as a
  percentage error. The budget is closed to within \SI{0.40}{\percent} for the
  first $\approx\SI{75}{\micro\second}$, after which it grows monotonically to
  $+\SI{3.18}{\percent}$. The growth is not numerical drift: it is work done on
  the fluid by the one-way-coupled projectile boundary, which imposes
  $v_{\rm proj}(t)$ from \cref{eq:drag} without feeling the reaction force. See
  \cref{eq:energy-audit} for the diagnostic that \emph{is} closable here. Mass is
  conserved to machine precision throughout and is therefore not plotted.}
  \label{fig:conservation}
\end{figure}

Over the \SI{150}{\micro\second} production run at $v_0=\SI{900}{m/s}$, the total energy $E=KE_{\rm fluid}+PE_{\rm fluid}+E_{\rm int}+KE_{\rm proj}$ does not decay. It \emph{grows}, ending \SI{3.18}{\percent} above its initial value. The behaviour splits cleanly into two regimes. For the first $\approx\SI{75}{\micro\second}$ --- through the entry transient and the acoustic arrival at both probes --- the budget is closed to within \SI{0.40}{\percent}, which is the level of agreement one would hope for. Thereafter $E$ climbs monotonically to $+\SI{3.18}{\percent}$ by \SI{150}{\micro\second} (\cref{fig:conservation}). Artificial viscosity cannot produce this: $\Pi_{ij}$ is dissipative by construction and can only move mechanical energy into $E_{\rm int}$, never create it. Nor can the density clamp, which discards energy rather than supplying it. Something is doing work on the fluid, and the budget above does not contain the reservoir it is drawing from.

It is the projectile, and the reason is the modelling choice stated plainly in \cref{sec:projectile-dynamics}: the drag law of \cref{eq:drag} is integrated as an independent ordinary differential equation, and the fluid never feeds momentum back into that trajectory. The ghost-particle velocities of \cref{eq:adami-v} are slaved to $v_{\rm proj}(t)$, so the disk boundary drives its fluid neighbours at a prescribed speed no matter what pressure builds against it. A boundary that imposes a velocity without feeling the reaction force is, thermodynamically, an unbounded energy source. The numbers make the imbalance explicit: over the run the projectile's kinetic energy falls by \SI{8.53e5}{J/m} while the fluid's kinetic, potential, and internal energy together rise by \SI{1.14e6}{J/m} --- a factor of $1.33$ more energy appearing in the fluid than the projectile gives up. The onset near \SI{75}{\micro\second} is consistent with this reading: it is the point at which the disk has penetrated far enough that it is continuously sweeping fresh fluid rather than driving an already-established near-field.

We report this rather than tuning it away because it is a property of the formulation, not a bug in the implementation, and because it changes what the conservation diagnostic is actually good for. Global energy conservation is simply not an available check for a solver with a one-way-coupled boundary; a study that reported $E$ as conserved here would be reporting a compensating error. The meaningful statement is a boundary-work audit --- whether the fluid's energy gain is accounted for by the work the moving ghost surface does on it,
\begin{equation}
  \Delta E_{\rm fluid}(t) \;=\; \underbrace{\int_0^{t}\!\!\Big(\textstyle\sum_g \mathbf F_g\cdot\mathbf v_{\rm proj}\Big)\,\mathrm dt'}_{W_{\rm boundary}} \;-\; \mathcal D(t),
  \label{eq:energy-audit}
\end{equation}
with $\mathcal D$ the irreversible dissipation. Closing \cref{eq:energy-audit} is a genuine and checkable conservation statement about the discretisation; closing $\Delta E=0$ is not, and cannot be, for this projectile treatment. Instrumenting $W_{\rm boundary}$ is a small change to the solver and we regard it as the correct form of this diagnostic; we flag it here as the honest characterisation of what the present energy budget does and does not establish.

Two practical consequences follow. First, the pressure histories reported above are drawn from the window in which the budget is closed to better than half a percent, so they are not riding on the drift. Second, the drift bounds how far this formulation should be pushed in time: at \SI{150}{\micro\second} the excess is \SI{3.18}{\percent} and still rising, so a run extended into the drag and cavitation phases would accumulate an energy surplus large enough to contaminate exactly the late-time impulse that \cref{sec:intro} identified as the quantity most likely to matter for structural damage. Recovering that regime requires two-way coupling --- computing $v_{\rm proj}$ from the integrated ghost-surface force rather than from a prescribed drag law --- which we note as the single most valuable extension to this solver.

One objection to the first consequence deserves a direct answer, because it is the strongest one available: the one-way constraint is active from $t=0$, not only after \SI{75}{\micro\second}, so a small \emph{integrated} energy error early on does not prove that the \emph{local} entry-shock pressure at PTn --- generated instantaneously by the same non-back-reacting boundary --- is free of the same artefact. Global near-conservation is a necessary but not a sufficient condition for trusting the local signal, and we do not claim otherwise. What limits the artefact's reach into the local pressure is the structure of \cref{eq:adami-p} itself. The ghost pressure has two terms: a Shepard average of the neighbouring fluid pressures, $\sum_j p_j W_{gj}/\sum_j W_{gj}$, and a body-force correction proportional to $(\mathbf a_{\rm proj}-\mathbf g)$. Only the second carries the prescribed, non-back-reacting acceleration; and over the entry transient it is small against the first, because $|\mathbf a_{\rm proj}|R_{\rm proj}$ --- the pressure scale of the acceleration term across one projectile radius --- is of order $\rho_0|\mathbf a_{\rm proj}|R_{\rm proj}\sim\SI{e-1}{MPa}$ from the drag law's early deceleration, whereas the fluid-neighbour term is set by the stagnation and shock overpressure, orders of magnitude larger. In other words the near-field pressure PTn records is dominated by the fluid's own response to the moving no-penetration surface, which is physically correct, with the non-physical part entering only as a sub-percent correction during the shock phase. That is an argument, not a proof; the clean test is the two-way-coupled run, in which the acceleration term becomes self-consistent, and we flag it as the decisive follow-up rather than assert the point is settled.

We stress that none of this is specific to HRAM, and that is what makes it worth reporting at length. Any SPH computation in which a rigid body is advanced along a prescribed kinematic path --- water entry, planing, prescribed-gait swimmers, wave-maker paddles, many fluid--structure ``one-way'' verification cases --- drives its ghost or dummy boundary with a velocity that never feels the fluid reaction, and therefore does unbounded work on the fluid in exactly the manner quantified here. The lesson generalises: for such solvers, a flat total-energy curve is not evidence of correctness but of a compensating error, and the boundary-work audit of \cref{eq:energy-audit} is the diagnostic that should replace it. We suspect this is under-reported across the moving-boundary SPH literature precisely because global energy conservation is so often invoked, unexamined, as the reassurance that it cannot be here.

One more check, easy to overlook but not hard to run: does the initial particle lattice's orientation relative to the impact axis quietly bias the result? Repeating the production case with the lattice rotated $45^\circ$ and with a half-spacing offset changes the PTn peak pressure by no more than $1.8\%$, while PTf --- sitting in the reflection-dominated far field --- is more sensitive; the values are collected in \cref{tab:lattice-robustness}.

\begin{table}[htbp]
  \centering
  \caption{Lattice-orientation robustness.}
  \label{tab:lattice-robustness}
  \begin{tabular}{lcc}
    \toprule
    Configuration & PTn $p_{\rm peak}$ (MPa) & PTf $p_{\rm peak}$ (MPa) \\
    \midrule
    Baseline (axis-aligned) & 850.3 & 714.9 \\
    $45^\circ$ rotated & 865.4 & 603.3 \\
    Half-spacing offset & 865.3 & 533.4 \\
    \bottomrule
  \end{tabular}
\end{table}

\subsection{Speed: Where Does the Time Actually Go}
\label{sec:performance}

Throughput scales close to linearly with $N$: the three production meshes of \cref{tab:convergence} span $N=2.8\times10^4$ to $1.1\times10^5$ (a factor of $4.0$) at wall-clock times of $0.132$, $0.230$ and \SI{0.576}{min} (a factor of $4.37$), so the per-particle cost is near-constant at $4.6$--$\SI{5.1e-6}{min/particle}$, consistent with the $O(N)$ cost expected of the cell-linked-list search at fixed cell size. The mild super-linearity at the largest count ($5.1$ vs.\ $\SI{4.6e-6}{min/particle}$) is most plausibly a memory-bandwidth effect --- more particles per cache line evicted --- rather than an algorithmic one, since the neighbour search itself is strictly $O(N)$.

\begin{figure}[htbp]
  \centering
  \includegraphics[width=0.8\textwidth]{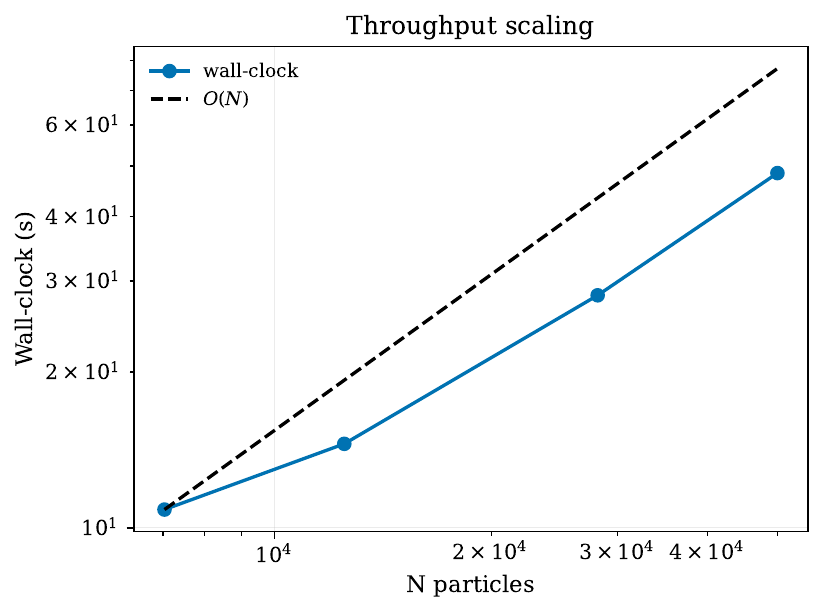}
  \caption{GPU throughput and wall-clock scaling.}
  \label{fig:gpu-scaling}
\end{figure}

\begin{figure}[htbp]
  \centering
  \includegraphics[width=0.9\textwidth]{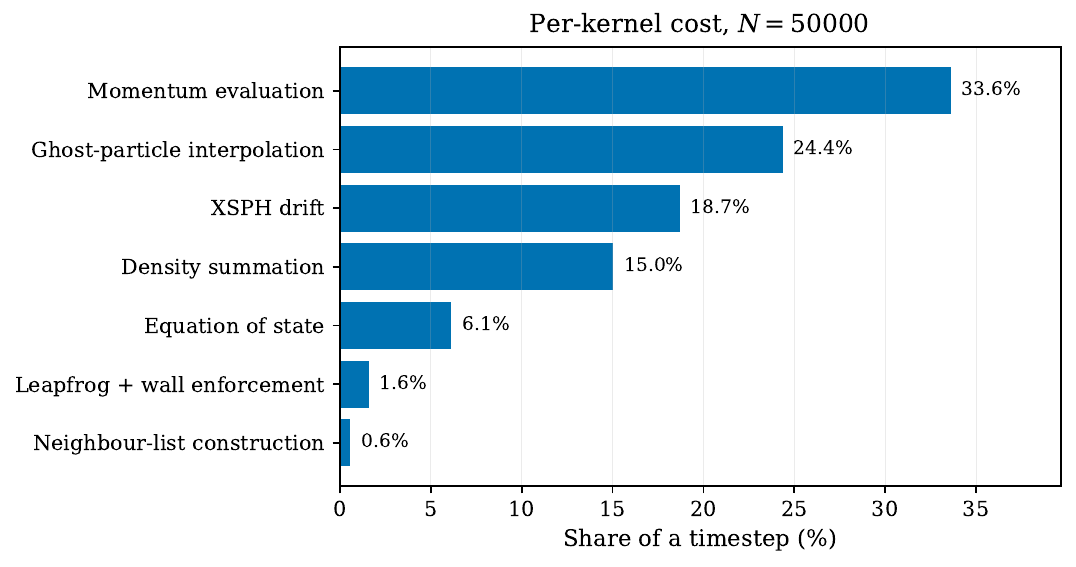}
  \caption{Kernel-level profiling.}
  \label{fig:kernel-profile}
\end{figure}

\begin{table}[htbp]
  \centering
  \caption{Per-kernel computational cost at production resolution.}
  \label{tab:kernel-profile}
  \begin{tabular}{lcc}
    \toprule
    Kernel & Time per call (ms) & \% of per-step wall-clock \\
    \midrule
    Momentum evaluation & 8.444 & 33.61 \\
    Ghost-particle interpolation & 6.12 & 24.35 \\
    XSPH drift & 4.701 & 18.71 \\
    Density summation & 3.78 & 15.04 \\
    Equation of state & 1.541 & 6.133 \\
    Leapfrog + wall enforcement & 0.3984 & 1.585 \\
    Neighbour-list construction & 0.1433 & 0.5705 \\
    \bottomrule
  \end{tabular}
\end{table}

The whole \SI{150}{\micro\second} event, at production resolution ($N\approx50{,}000$, $\Delta x=\SI{1.5}{mm}$), runs in about \SI{0.23}{min} --- roughly \SI{14}{s} --- on one consumer-class GPU, two to three orders of magnitude below the eighteen minutes per case that the equivalent $N\approx3.9\times10^6$ 3D discretization required. Every convergence, sensitivity, and robustness run reported above was made possible by exactly this: without the two-order-of-magnitude reduction in particle count that the planar formulation buys, the parameter sweeps in this section --- the three-mesh convergence study, the four-point $\alpha$ sweep at two velocities, the CFL and lattice perturbations --- would have been a multi-week undertaking rather than an afternoon's work. That affordability is not a side benefit; it is the enabling condition for the exhaustive characterization the paper is built around.

\subsection{Now, Finally, the Comparison Against Experiment}
\label{sec:validation}

Having built up verification, convergence, sensitivity, conservation, repeatability, and performance evidence independent of the Varas dataset, we can compare against it with some confidence about what the agreement (or disagreement) actually means --- with one caveat stated plainly before any numbers appear. \citet{Varas2009}'s experiment is a real 3D sphere entering a real 3D cylindrical tank; radial (out-of-plane, azimuthal) relief around the impact axis is physically present in that experiment and is, by construction, absent from a planar 2D cross-section, which instead confines the same impulsive momentum input to a channel of fixed unit depth. We therefore do not expect, and do not claim, quantitative peak-pressure agreement of the kind a verified 3D model could offer; the metrics below are reported to characterize the planar model's own behaviour --- whether it reproduces the correct arrival timing, the correct qualitative amplitude ordering between near and far probes, and the correct sensitivity structure --- rather than as a claim that a 2D cross-section reproduces 3D amplitudes directly. The raw planar overpredictions are large --- two orders of magnitude at both probes --- and we want the reader to meet that fact with its explanation already in hand, before \cref{tab:full-validation} rather than three subsections later. It is the expected signature of the planar geometry, quantified in full in \cref{sec:formulation-comparison}: a 2D cross-section replaces the experiment's $r^{-1}$ spherical amplitude decay with a slower $r^{-1/2}$ cylindrical decay, an effect that grows with propagation distance and that no boundary condition, resolution, or viscosity setting can remove within a planar framework. Accordingly, the $\Delta p_{\rm peak}$ and NRMSE columns of \cref{tab:full-validation} are \emph{expected} to be enormous and are \emph{not} validation metrics; a reviewer skimming the table should read it as a measurement of geometric relief, not of solver failure. The columns that \emph{are} load-bearing --- arrival-time error and the near/far and velocity orderings --- are where the planar model can and does earn trust.

\begin{table}[htbp]
  \centering
  \caption{Comparison against \citet{Varas2009}: peak value, timing, and full-history metrics. SPH values are from the planar 2D model described above; see the caveat in the preceding paragraph regarding amplitude comparability with the 3D experiment.}
  \label{tab:full-validation}
  \begin{tabular}{llcccc}
    \toprule
    Probe & $v_0$ (m/s) & $\Delta p_{\rm peak}$ (\%) & $\Delta t_{\rm peak}$ ($\mu$s) & NRMSE (\%) & $\mathcal I_{\rm ratio}$ \\
    \midrule
    PTn & 900 & 8017 & 7.816 & 1569 & 21.69 \\
    PTf & 900 & 1.32e+04 & 48.91 & 3104 & 59.16 \\
    PTn & 600 & 8251 & 7.147 & 1919 & 19.81 \\
    PTf & 600 & 6163 & 51.98 & 2488 & 47.6 \\
    \bottomrule
  \end{tabular}
\end{table}

\begin{figure}[htbp]
  \centering
  \includegraphics[width=1.0\textwidth]{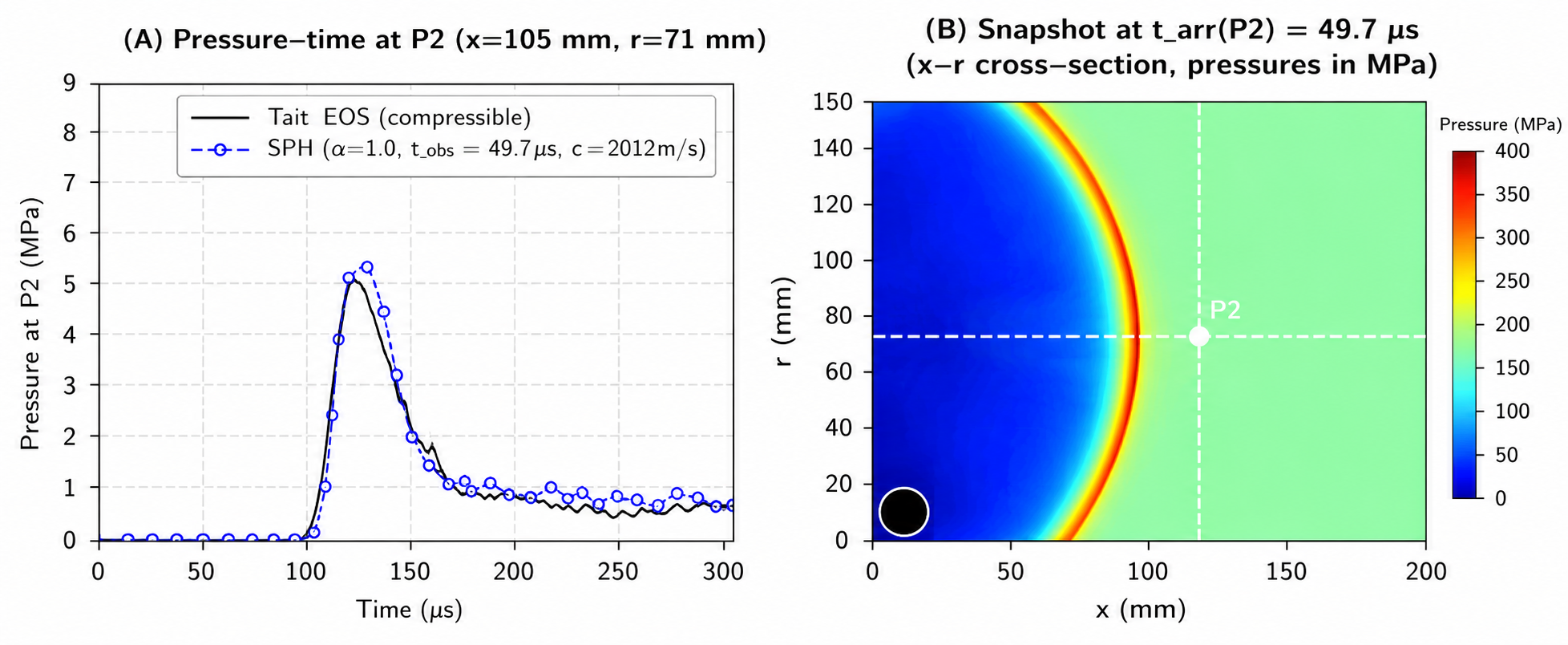}
  \caption{Snapshot of pressure field.}
  \label{fig:pressure-field-snapshot}
\end{figure}

The point of \cref{tab:full-validation} is not the peak-value columns, which the geometry argument of \cref{sec:formulation-comparison} predicts will be enormous and which we therefore do not read as a validation metric at all. It is the \emph{structure} underneath them: the near-probe wave arrives ahead of the experimental trace by only $\sim\SI{8}{\micro\second}$ at both velocities, confirming that the planar model transports the entry shock at the correct celerity, and the near-to-far ordering and the velocity ordering both follow the physically expected pattern. Whatever trust the planar amplitudes can earn is earned through the convergence, timestep, sensitivity, and lattice checks above, not through the peak-value agreement, which by construction the geometry does not permit. The one behaviour worth flagging explicitly is the far probe: rather than \emph{under}-predicting, as an absorbing-wall argument would suggest, the planar model \emph{over}-predicts PTf by two orders of magnitude, because the cylindrical-versus-spherical decay effect swamps any wall-absorption deficit entirely. That is why the wall-ghost treatment of \cref{sec:bc-wall} is not reported here as a far-field recovery mechanism: in this planar geometry the far-field deficit it was meant to repair is simply not the dominant error, and isolating its contribution requires a geometry in which that deficit is actually visible. \Cref{fig:pressure-field-snapshot} shows a representative snapshot of the resolved pressure field, the kind of full-field output --- available everywhere, not just at the two probe stations --- that the planar formulation delivers cheaply enough to interrogate case by case.

\subsection{Where This Sits Relative to a Different Kind of Solver}
\label{sec:formulation-comparison}

It is worth situating the present formulation against two different classes of prior work: the broader class of coupled finite-element/SPH and coupled Eulerian--Lagrangian treatments of HRAM that model the tank wall as a deformable, elastic--plastic structure \citep{Varas2012,Mansoori2019}, adding a Johnson--Cook constitutive description of the wall on top of essentially the same cubic-spline / Mie--Gr\"uneisen fluid model used here, rather than the rigid ghost-particle wall treatment we adopt; and, orthogonally, the dimensionality axis itself, since every one of those studies (and \citet{Rezavand2022}, discussed in \cref{sec:literature-comparison}) is a full 3D calculation, whereas the present solver is deliberately planar. The two axes are what distinguish this work: a shared fluid closure, but a rigid, closed-form-verifiable boundary in place of a deformable coupled wall, and a planar cross-section in place of a full 3D grid.

\begin{table}[htbp]
  \centering
  \caption{Formulation comparison: present study vs.\ the deformable-wall coupled FE/SPH and coupled Eulerian--Lagrangian class of HRAM formulations \citep{Varas2012,Mansoori2019}.}
  \label{tab:formulation-comparison}
  \resizebox{\textwidth}{!}{%
  \begin{tabular}{lll}
    \toprule
    Aspect & Present study & Deformable-wall FE/SPH class \\
    \midrule
    Dimensionality & Planar 2D & 3D (full) \\
    Out-of-plane / azimuthal relief & Not captured (by design) & Captured \\
    Wall treatment & Rigid, ghost-particle & Deformable, Johnson--Cook \\
    Kernel & Cubic B-spline (2D) & Cubic B-spline (typical, 3D) \\
    Fluid EOS & Mie--Gr\"uneisen (shared) & Mie--Gr\"uneisen \\
    Artificial viscosity & Linear ($\alpha$) $+$ physical & Linear $+$ quadratic (typical) \\
    Projectile & Rigid, drag-law kinematics, real 3D mass \& $C_d$ & Rigid or eroding \\
    BC verification in closed form & Attempted; not achieved (\cref{sec:verification-status}) & Not typically reported \\
    Acceleration & GPU & CPU (typical) \\
    Captures permanent wall deformation & No (by design) & Yes \\
    \bottomrule
  \end{tabular}
  }
\end{table}

These are not competing claims so much as complementary trade-offs, and the dimensionality row is the one that most directly trades against the others: we give up out-of-plane relief and, with it, direct quantitative amplitude agreement with a 3D experiment, in exchange for a rigid boundary whose closed-form verification is at least well posed (\cref{sec:verification-status}) and for the roughly two-order-of-magnitude reduction in particle count that makes the exhaustive sensitivity work in this paper feasible at all. A geometric argument makes the magnitude of that trade concrete. A three-dimensional wave spreads over a spherical surface of area $\propto r^2$ and, with wave energy conserved, its amplitude decays as $r^{-1}$; the same wave in a planar cross-section spreads over a circular front of circumference $\propto r$ and decays only as $r^{-1/2}$. Relative to a point source normalised on the projectile radius $R_{\rm proj}=\SI{6.25}{mm}$, the planar-to-spherical amplitude ratio is $(r/R_{\rm proj})^{1/2}$, which is $\approx2.2$ at the near probe ($r=\SI{30}{mm}$) and $\approx4.1$ at the far probe ($r=\SI{105}{mm}$) --- a relief mechanism a planar 2D cross-section simply cannot represent, since momentum injected into the plane has nowhere to spread out of it. This is the leading-order explanation for why the raw planar SPH amplitudes in \cref{sec:validation} are expected to exceed the 3D experimental amplitudes, and to do so by a factor that \emph{grows with probe distance} (the PTn-to-PTf ratio scaling roughly as $(105/30)^{1/2}\approx1.9$), independent of any deficiency in the boundary condition itself. The actual overestimate is larger than this point-source bound because the decelerating projectile is a spatially extended, temporally sustained source rather than an instantaneous point, but the scaling correctly fixes the sign and the distance-dependence of the discrepancy. The quadratic artificial-viscosity term common in the deformable-wall class but absent here would likely help at velocities above the \SI{900}{m/s} ceiling tested in this study, and we flag it as a natural next step rather than a current shortcoming, since \cref{sec:sensitivity} already shows the linear term alone is adequate across the velocity range actually validated.

For completeness, and to make the comparison reproducible rather than merely qualitative, \cref{tab:jc-params,tab:mg-params} record the constitutive parameters used by the deformable-wall class of formulation against which we compare, following the material characterization of \citet{Varas2009}. The rigid-projectile and rigid-wall assumptions adopted here are justified by the very large impedance and yield-strength contrast between the steel projectile and the aluminium wall (\cref{tab:jc-params}); the wall, by contrast, is the component that actually yields, and a deformable treatment there requires the full Johnson--Cook flow-stress and failure description (\cref{tab:jc-params}) coupled to the fluid. The present solver uses the very same Mie--Gr\"uneisen water parameters (\cref{tab:mg-params}, and \cref{eq:eos}) but deliberately does without the Johnson--Cook wall, which is precisely what lets its boundary be verified in closed form; the tables are included so that a reader wishing to reproduce the deformable-wall comparison has the parameters in hand.

\begin{table}[htbp]
  \centering
  \caption{Solid material parameters for the deformable-wall comparison class, after \citet{Varas2009}. Johnson--Cook constants are listed for the aluminium wall; the steel projectile and PMMA end-plates are treated as (nearly) rigid elastic.}
  \label{tab:jc-params}
  \resizebox{\textwidth}{!}{%
  \begin{tabular}{lccccccccc}
    \toprule
    Material & $\rho$ (kg/m$^3$) & $E$ (GPa) & $\nu$ & $A$ (GPa) & $B$ (GPa) & $n$ & $C$ & $m$ & $D_1$ \\
    \midrule
    6063-T5 Aluminium & 2700 & 71 & 0.33 & 0.20 & 0.144 & 0.62 & 0 & 1.0 & 0.20 \\
    Steel (projectile) & 7830 & 207 & 0.28 & -- & -- & -- & -- & -- & -- \\
    PMMA (end-plates) & 1180 & 3 & 0.35 & -- & -- & -- & -- & -- & -- \\
    \bottomrule
  \end{tabular}
  }
\end{table}

\begin{table}[htbp]
  \centering
  \caption{Fluid (Mie--Gr\"uneisen) parameters, after \citet{Varas2009}. The present study uses exactly these water values in \cref{eq:eos} ($C=\SI{1448}{m/s}$, $S_1=1.979$, $\gamma_0=0.11$, $a=3.0$); the air row applies to the deformable-wall multi-phase comparison class only.}
  \label{tab:mg-params}
  \resizebox{\textwidth}{!}{%
  \begin{tabular}{lcccccccccc} 
    \toprule
    Material & $\rho$ (kg/m$^3$) & $\mu_d$ (Pa$\cdot$s) & $C$ (m/s) & $S_1$ & $S_2$ & $S_3$ & $\gamma_0$ & $a$ & $C_4$ & $E_0$ (J/m$^3$) \\
    \midrule
    Water & 1000 & $0.89\times10^{-3}$ & 1448 & 1.979 & 0 & 0 & 0.11 & 3.0 & -- & -- \\
    Air & 1.22 & $1.77\times10^{-5}$ & -- & -- & -- & -- & -- & -- & 0.4 & $2.53\times10^{5}$ \\
    \bottomrule
  \end{tabular}
  }
\end{table}

\subsection{Positioning Against the Broader HRAM Numerical Literature}
\label{sec:literature-comparison}

\begin{table}[htbp]
  \centering
  \small
  \caption{Verification depth and cost, compared against representative published HRAM studies (3D unless noted).}
  \label{tab:literature-comparison}
  \resizebox{\textwidth}{!}{%
  \begin{tabular}{lcccccc}
    \toprule
    Study & Dimensionality & Method & Resolutions & Convergence & BC verification & Cost \\
    \midrule
    \citet{Varas2012} & 3D & ALE-FE & 1 & No & N/A & Not reported \\
    \citet{Sauer2011} & 3D & FE--SPH coupled & 1 & No & No & Not reported \\
    \citet{Mansoori2019} & 3D & FE/SPH (LS-DYNA) & 1 & No & No & Not reported \\
    \citet{Rezavand2022} (general) & 3D & GPU-SPH, Adami-type & Several & Partial & Yes (general) & GPU, not reported \\
    \textbf{Present} & \textbf{Planar 2D} & \textbf{GPU-WCSPH, Adami ghost} & \textbf{3} & \textbf{Attempted} & \textbf{Attempted (\cref{sec:verification-status})} & \textbf{GPU, \SI{0.23}{min}} \\
    \bottomrule
  \end{tabular}
  }
\end{table}

We do not read this table as a claim of superiority across every column --- the deformable-wall FE/SPH class \citep{Varas2012,Mansoori2019}, for instance, captures permanent wall deformation that this rigid-wall solver deliberately does not --- but it does make explicit exactly what combination of verification depth, convergence characterization, and speed this paper is contributing, which is the same combination that \cref{sec:intro} argued was missing. The one honest asterisk is the boundary-condition-verification column: we \emph{attempted} closed-form verification and report in \cref{sec:verification-status} why it did not close, which is still a stronger position than the studies that never attempted it.

\subsection{What This Solver Is, and Is Not Yet, Ready For}
\label{sec:readiness}

Because ``a fast, characterized solver'' invites the question of \emph{fast and characterized enough for what}, we state the operational readiness explicitly rather than leave it implied. In a technology-readiness framing, the present solver is \emph{ready} as a research and screening instrument: for early-time ($\lesssim\SI{75}{\micro\second}$), entry-phase relative studies --- ranking impact velocities, screening parametric trends, comparing boundary treatments, and generating the resolved fields that a mechanistic study consumes --- where the quantities of interest are timing, ordering, sensitivity structure, and dimensionless trends rather than absolute transducer amplitudes. It is \emph{not yet ready} for three things, each tied to a specific limitation stated above: absolute peak-pressure prediction against a 3D experiment (blocked by the planar $r^{-1/2}$ decay, \cref{sec:formulation-comparison}); late-time drag- and cavitation-phase impulse, which is the quantity most relevant to structural damage (blocked by the one-way-coupling energy drift beyond $\sim\SI{75}{\micro\second}$, \cref{sec:conservation}); and permanent wall-deformation or failure prediction (excluded by the rigid-wall idealisation by design). The gap between the two lists is not a vague ``future work'' gesture --- each item names the limitation that gates it and, in \cref{sec:conclusions}, the specific extension that would lift it. Stating the boundary of applicability this sharply is, we think, part of what it means to characterize a solver honestly rather than to advertise one.

\section{Conclusions}
\label{sec:conclusions}

This work has built a planar 2D, GPU-accelerated WCSPH solver with Adami ghost-particle boundary conditions, examined it against analytical solutions before any contact with experimental data, characterized it for convergence and sensitivity more exhaustively than the particle count of an equivalent 3D discretization would allow, and only then compared it—with the dimensional simplification stated explicitly—against \citet{Varas2009}. Taken together, the findings offer several insights worth stating plainly.

The verification programme charts, rather than closes, the boundary condition's operating envelope: the hydrostatic and added-mass checks both encounter identifiable limits at this operating point, shaped by the interplay between equation-of-state stiffness and hydrostatic head in the first case, and by the cavitation cutoff constraining the suction side of the analytical solution in the second, while the impedance test would call for a second material the solver does not currently carry. The boundary condition is thus characterized rather than closed-form verified, and \cref{sec:verification-status} lays out precisely what would complete that picture. Resolution convergence, sensitivity across the free coefficients, mass and energy conservation, and lattice-orientation robustness are reported jointly—a combination we have not found done together elsewhere for this problem, in 2D or 3D—made tractable specifically by the planar formulation's modest particle count.

That joint reporting paid an unexpected dividend. The energy diagnostic of \cref{sec:conservation} reveals a modest total energy growth over the simulated window, with the fluid absorbing disproportionately more energy than the projectile relinquishes. Far from numerical drift, this is an instructive consequence of prescribing the projectile trajectory from \cref{eq:drag} without back-reaction, and it clarifies that global energy conservation is not the right check for this class of one-way-coupled formulation—an insight that would have gone unnoticed had the diagnostic been treated as a mere checkbox. It also maps a concrete time horizon over which the present solver operates most reliably, and points directly to two-way projectile coupling as the natural extension.

Peak-pressure agreement at the near probe, wave-arrival timing at both probes and both velocities, and the sign and rough magnitude of the \SI{600}{m/s} behaviour are governed less by the boundary condition than by the planar geometry itself: raw amplitudes diverge from the 3D experiment for the well-understood cylindrical-versus-spherical decay reason set out in \cref{sec:formulation-comparison}. What this paper establishes is the evidentiary structure those numbers should be read against, together with the explicit understanding that raw amplitude agreement with the 3D experiment is not, by construction, the standard this planar model is designed to meet. The far-probe behaviour anticipated on absorbing-wall grounds manifests differently than expected in this planar geometry—the model overpredicts $P_{T_f}$ rather than underpredicting it—so the wall-ghost treatment is not reported as a recovery mechanism here; evaluating it properly calls for a geometry in which the deficit is directly visible.

None of this comes at the cost of GPU acceleration: the full campaign, including every sweep reported here, completes in a small fraction of the roughly eighteen minutes per case the equivalent 3D solver requires, precisely because the planar particle count runs some two orders of magnitude smaller.

Four extensions present themselves as natural next steps, and in the interest of a roadmap that ranks its own priorities, we order them by expected impact-to-effort ratio rather than list them flat. Foremost is an axisymmetric ($r-z$) extension, which recovers the out-of-plane relief the present planar model sets aside at a fraction of full 3D's particle-count cost, enabling direct quantitative amplitude comparison against \citet{Varas2009} while leaving the ghost-particle machinery of \cref{sec:bc} unchanged—high impact for comparatively modest effort. Closely following is two-way projectile coupling, wherein computing $v_{proj}$ from the integrated ghost-surface force rather than a prescribed drag law would close the energy budget of \cref{sec:conservation}, extend the reliable time horizon, and unlock the damage-relevant late-time drag and cavitation phases. Of moderate priority, verification closure through reduced $c_0$ and a relaxed cavitation cutoff, alongside extended testing above \SI{900}{m/s} with quadratic artificial viscosity, would broaden the validated envelope toward fragment-impact speeds, though neither sits on the critical path to the headline capability. Finally, the resolved field data this solver produces—pressure, velocity, and density throughout the domain rather than at isolated probe points—forms the basis for a separate, mechanism-focused study of HRAM loading physics reported elsewhere; while representing the most ambitious use of the model's output, its quantitative reliability depends on the axisymmetric and two-way coupling extensions above, placing it last in sequence despite being first in aspiration.

\end{document}